\documentclass[floatfix,aps,twocolumn,preprintnumbers,amsmath,amssymb,nofootinbib,superscriptaddress,showkeys]{revtex4}

\usepackage{graphicx,color}
\usepackage{amsmath,amssymb,bm}
\usepackage{dcolumn}
\usepackage{comment}

\newcommand{\bea}{\begin{eqnarray}}
\newcommand{\eea}{\end{eqnarray}}
\newcommand{\be}{\begin{equation}}
\newcommand{\ee}{\end{equation}}
\newcommand{\np}{{\bf p}}

\newcommand{\nr}{{\bf r}}

\newcommand{\nh}{{\bf h}}

\newcommand{\nq}{{\bf q}}

\def\XXint#1#2#3{{\setbox0=\hbox{$#1{#2#3}{\int}$}
     \vcenter{\hbox{$#2#3$}}\kern-.5\wd0}}

\def\1{\'{\i}}

\begin{document}

\title{
Extended superscaling with two-particle emission 
in electron and neutrino scattering
}
\author{V.L. Martinez-Consentino}\email{victormc@ugr.es} 
  \affiliation{Departamento de
  F\'{\i}sica At\'omica, Molecular y Nuclear \\ and Instituto Carlos I
  de F{\'\i}sica Te\'orica y Computacional \\ Universidad de Granada,
  E-18071 Granada, Spain.}

\author{J.E. Amaro}\email{amaro@ugr.es} \affiliation{Departamento de
  F\'{\i}sica At\'omica, Molecular y Nuclear \\ and Instituto Carlos I
  de F{\'\i}sica Te\'orica y Computacional \\ Universidad de Granada,
  E-18071 Granada, Spain.}
  
\author{P.R. Casale} \affiliation{Departamento de
  F\'{\i}sica At\'omica, Molecular y Nuclear \\ and Instituto Carlos I
  de F{\'\i}sica Te\'orica y Computacional \\ Universidad de Granada,
  E-18071 Granada, Spain.}

\author{I. Ruiz Simo}\email{ruizsig@ugr.es} \affiliation{Departamento de
  F\'{\i}sica At\'omica, Molecular y Nuclear \\ and Instituto Carlos I
  de F{\'\i}sica Te\'orica y Computacional \\ Universidad de Granada,
  E-18071 Granada, Spain.}

\date{\today}

\begin{abstract}
An extended superscaling analysis of quasielastic electron scattering
data is proposed by parametrizing the scaling function as the sum of a
symmetric function corresponding to the emission of a single particle
plus a contribution from the phase space of two-particle emission.
The phase space of two-particle emission (2p2h) is multiplied by a
q-dependent parameter that has been fitted to describe the
tail behavior of the scaling function. This approach allows for an
alternative description of the quasielastic electron scattering
data, incorporating the contributions from both single-particle and
two-particle emission processes induced by the one-body current
and explaining the asymmetry of the scaling function.
In a factorized schematic model based on the independent-pair
approximation, the 2p2h parameter is related to the high-momentum
distribution of the pair averaged over 2p-2h excitations. However, in the
phenomenological fitting approach undertaken here, this coefficient
includes other contributions such as interference with two-body
currents and effects of the final-state interactions.  We
present predictions for the inclusive two-nucleon emission cross
section induced by electrons and neutrinos.

\end{abstract}


\keywords{
Quasielastic electron scattering,
neutrino scattering,
Two-particle emission
 }

\maketitle

\section{Introduction}

Knowledge of the neutrino-nucleus interaction and the
electron-nucleus interaction is essential for current neutrino
experiments with accelerators
\cite{Ank22,Ama20,Mos16,Kat17,Alv14,Ank17,Ben17}.  
The
emission of one nucleon is the most important contribution to the
inclusive cross section in the quasi-elastic (QE) region, centered around
$\omega=|Q^2|/2m^*_N$, where $\omega$ is the energy transfer,
$Q^2=\omega^2-q^2<0$, and $q$ is the momentum transfer to a nucleon
with relativistic effective mass $m_N^*$
\cite{Ros80,Ser86,Dre89,Weh93}.  Recent theoretical work have
shown the importance of two-particle (2p2h) excitation in the
QE cross section (about 20\% of the total cross section)
\cite{Mar09,Nie11,Gal16,Meg14,Meg16,Meg16b,Ank15,Gra13,Pan16,Mar16,Mar21,Mar21b}
.

The emission of two particles requires interaction mechanisms with a
pair of nucleons. Regardless of the final-state interactions, this can
be achieved with meson-exchange currents (MEC) \cite{Pac03,Rui17} and
with short-range correlations (SRC) models
\cite{Cuy16,Ryc97,Ste18,Cos21}.  Alternatively SRC have also been
introduced as a two-body correlation operator
\cite{Mar09,Alb84,Alb90,Ama02,Ama10}.  In ref. \cite{Nie11} the
two-body one-pion exchange operator is modified to include SRC via the
Landau Migdal parameters.  One effect of SRC is to provide high
momentum components to nucleons due to the nuclear force at short
distances \cite{Ste18,Ryc19,Cos21,Sim17}.  This is being exploited to
extract the number of SRC pairs from semi-inclusive and inclusive
$(e,e'NN)$ reactions \cite{Col15,Wei21,Ngu20}.  However, in general,
SRC and MEC
interfere with each other, and it is not possible to separate or
isolate them from other effects such as final-state interactions.

In this work, we present a method to obtain the contribution to the
inclusive response of 2p2h (two-particle, two-hole) induced by the
one-body current from the phenomenological scaling function.  The QE
scaling function is obtained from the $(e,e')$ data by dividing by a
single nucleon cross section \cite{Ama17,Mar17,Ama18}.  The starting
point is the superscaling analysis with relativistic effective mass
(SuSAM*) that is based in the relativistic mean field model (RMF) of
nuclear matter \cite{Ros80,Ser86}.  The SuSAM* scaling function
describes most of the QE events but it contains residual effects due
to SRC, MEC and other contributions.  In Ref. \cite{Mar21} the MEC
2p2h contribution was subtracted from the $(e,e')$ data, using a RMF
calculation, and a new scaling function was obtained without
contamination from the MEC 2p2h channel.  But this function still
contains contributions from other non-QE mechanisms, in particular
from 2p2h excitations produced by the one-body current, including SRC,
interferences with MEC, and final state interactions (FSI).

To extract the 2p2h contribution of the one-body (OB) current we assume that the
tail of the scaling function, $f^*(\psi^*)$, for high values of the
scaling variable, $\psi^*$ ---or equivalently for high values of the
energy transfer--- is due mainly to the 2p2h phase space.  In
particular SRC produce high momentum components in the nuclear wave
function while the excitation of nucleons with high momentum implies
high values of the scaling variable, $\psi^*> 1$, which gives a
contribution to the tail of $f^*(\psi^*)$.  It is important to note
that the contribution of SRC to the 2p2h
response is entangled with other effects, such as interference with
MEC and FSI
effects. This entanglement makes it not possible to extract the SRC
contribution unambiguously from the data alone.  In this work we
assume that the 2p2h response function induced by OB current can be
parametrized with a semiempirical formula similar to that of the MEC
responses \cite{Mar21b}. By this assumption the 2p2h response will
be proportional to the 2p2h phase space and to the single-nucleon
response. The combined effects of SRC, MEC, FSI, and possibly
other nuclear dynamics will be encoded in 2p2h parameters fitted to
reproduce the tail of the scaling function.  Our approach is
alternative and complementary to current search on SRC pairs from
inclusive electron scattering data in the zone of low energy and high
transferred momentum $q>1.5$ GeV/c.  \cite{Wei21}.

In this work we focus on intermediate momentum transfer $q\leq 1$
GeV/c.  This region is of interest for neutrino experiments where we
are going to apply our results.  One of the processes we will analyze
in this work is the emission of two particles as a result of energy
and momentum transfers to a pair of correlated nucleons whose wave
function contains high-momentum components.  In the independent pair
approximation, the high-momentum component is thought to originate
from the interaction between nucleon pairs in the nuclear medium,
which can be mathematically expressed as the solution to the
Bethe-Goldstone (BG) equation.  We will demonstrate that the product
of the OB current with the high-momentum wave function of a nucleon
pair leads to an effective two-body correlation current.  If we make
the approximation of factorizing an average of the one-body current
and the phase space in this model, we obtain a coefficient that is
related to the average of the high-momentum distribution of a pair of
correlated nucleons. 

In our scaling analysis, we observe that the scaled data, when plotted
as a function of the scaling variable, exhibit a compatible behavior
and can be effectively parameterized as the sum of a symmetric scaling
function and a 2p2h-like contribution. This 2p2h-like contribution,
proportional to the 2p2h phase space, accurately reproduces the tail
of the data for high values of $\omega$,
the 2p2h coefficients now being adjustable parameters to describe the
tail of the scaling function.

To obtain the scaled data, we divide the cross section by the averaged
single nucleon response. By doing so, we can assume that the 2p2h
component of the scaling function arises from the 2p2h emission
induced by the one-body current. This assumption is based on the
observation that the 2p2h-like contribution captures the high-energy
tail of the data.  By parametrizing the scaling function in this way,
we are able to disentangle the 2p2h contribution from the 1p1h one
and extract information about the underlying dynamics of the
reaction. This allows us to investigate the role of 2p2h processes
induced by the one-body current and their impact on the scaling
behavior observed in the experimental data.

In the SuSAM* approach the superscaling analysis of electron
scattering data is used to predict QE neutrino-nucleus
cross sections \cite{Rui18}. The neutrino (antineutrino) cross section
was extended to the 2p2h sector with a relativistic MEC operator in
the RMF model of nuclear matter \cite{Mar21b}, where the semiempirical
formula for two-nucleon emission responses was fitted to the exact
results for momenta in the range $q=200$ and 2000 MeV/c.  The
semiempirical formula allows to compute accurately the 2p2h MEC
responses using an analytical formula, thus
reducing the calculation time. 
Similarly, in this work, we extend the new parametrization of the
scaling function to the case of neutrino scattering. We use the same
2p2h parameters obtained from electron scattering, but multiply them
by the weak responses of a nucleon instead of the electromagnetic
responses. By applying this approach to neutrino scattering, we aim to
investigate the role of 2p2h processes induced by the weak one-body
current and their contribution to the  charged-current (CC) neutrino cross section.  By
utilizing the same 2p2h parameters, we can effectively compare the
2p2h contributions in electron and neutrino scattering and assess the
universality of the scaling behavior across different interaction
channels.

The scheme of the paper is as follows.  In Sect. II we present the
superscaling formalism. In sect. III we study the particular case
of the correlation current in the independent pair approximation.  
In sect. IV we introduce the semiempirical formula for the 2p2h response
function induced by the OB current. In sect. V we presents the results.
Finally,  in Sect. VI we draw our conclusions.

\section{Formalism}

We follow the formalism of Ref. \cite{Ama20}.  
We start with the inclusive electron scattering 
cross section in plane-wave Born approximation 
with one photon-exchange
\begin{equation}  \label{cross}
\frac{d\sigma}{d\Omega d\epsilon'}
= \sigma_{\rm Mott}
(v_L R_L(q,\omega) +  v_T  R_T(q,\omega)),
\end{equation}
where the incident electron has energy $\epsilon$, the scattering
angle is $\theta$, the final energy is energy $\epsilon'$, and
$\Omega$ is the solid angle for the electron detection.  The energy
transfer is $\omega=\epsilon-\epsilon'$, the momentum transfer is
$\nq$.
In Eq. (\ref{cross}) $\sigma_{\rm Mott}$ is the Mott cross section, 
$v_L$ and $v_T$ are
kinematic factors coming from the leptonic tensor
\begin{equation}
v_L = 
\frac{Q^4}{q^4}, 
\kern 1cm
v_T =  
\tan^2\frac{\theta}{2}-\frac{Q^2}{2q^2}.
\end{equation}
Finally, the longitudinal and transverse response functions, 
only depend on $(q,\omega)$ and 
are the following components of the hadronic tensor
\begin{eqnarray}
R_L(q,\omega) =  W^{00},
\kern 1cm
R_T(q,\omega)  =  W^{11}+W^{22},
\end{eqnarray}
where $W^{\mu\nu}$ is the nuclear hadronic tensor, see
Ref. \cite{Ama20} for details.

In refs. \cite{Mar21,Mar21b} we developed a model of the
reaction where the responses are the sum of QE plus 2p2h
MEC contribution
\begin{equation}
R_K(q,\omega)=R_K^{QE}(q,\omega) + R_K^{MEC}(q,\omega) + \cdots,
\kern 1mm 
\end{equation}
for $K=L,T$, where contributions of pion emission,
and beyond are not included in our model.  The MEC responses
were studied in detail in \cite{Mar21b}, where a semiempirical
formula was derived from a microscopical relativistic current
in the RMF model of nuclear matter.  We focus in this work on the QE
responses that we describe using the SuSAM* approach
\cite{Ama18,Mar21}.  This model is an extension of the RMF model of
nuclear matter \cite{Ros80}, where 
the initial and final nucleons 
are interacting with the nuclear mean field
and acquire an effective mass $m_N^*$.  
The 1p1h response functions
are
 \begin{eqnarray}
R_K^{QE}(q,\omega)
&=& 
 \frac{V}{(2 \pi)^3}
\int
d^3h
\frac{(m^*_N)^2}{EE'}
\delta(E'-E-\omega)
\nonumber\\
&& \times
\theta(p'-k_F)
\theta(k_F-h)  2U_K.  \label{rmf}
\end{eqnarray}
The
initial nucleon has momentum $\nh$ below the Fermi momentum, $h <
k_F$,  and on-shell energy $E=\sqrt{\nh^2+m_N^*{}^2}$.  The final
nucleon has momentum $\np'=\nh+\nq$, and the final on-shell energy is
$E'=\sqrt{\np'{}^2+m_N^*{}^2}$.  Pauli blocking implies $p' > k_F$.
The $U_K$ are the single-nucleon responses for the 1p1h excitation
\begin{eqnarray}
U_L =  w^{00},
\kern 1cm
U_T  =  w^{11}+w^{22},
\end{eqnarray}
corresponding to the single-nucleon hadronic tensor
\begin{equation}
w^{\mu\nu}=\frac12 \sum_{ss'} j^{\mu}_{OB}(\np',\nh)_{s's}^* \,
j^{\nu}_{OB}(\np',\nh)_{s's}
\end{equation}
and $j_{OB}^\mu$ is the electromagnetic current matrix element
\begin{equation} \label{job}
j^{\mu}_{OB}(\np',\nh)_{s's}= \overline{u}(\np')_{s'}
\left[F_1\gamma^\mu+i\frac{F_2}{2m_N}\sigma^{\mu\nu}Q_{\nu}
\right]u(\nh)_s,
\end{equation}
where $F_1$ and $F_2$, are the Dirac and Pauli form factors.
To compute the integral of Eq. (\ref{rmf}), the usual procedure is to
change from variable $\theta_h$ to $E'$. The integral over $E'$ is made using
 the
Dirac delta, this fixes the value of the angle between $\nh$ and $\nq$,
$\cos\theta_h= (2E\omega+Q^2)/(2hq)$, 
 and the integration over the azymuthal angle $\phi$ gives
$2\pi$ by symmetry of the responses when $\nq$ is on the $z$-axis
\cite{Ama20}. We are left with an integral over the initial nucleon
energy
 \begin{equation}
R_K^{QE}(q,\omega)
= 
 \frac{V}{(2 \pi)^3}
 \frac{2\pi m_N^{*3}}{q}
\int_{\epsilon_0}^{\infty}d\epsilon\, n(\epsilon)\, 2U_K(\epsilon,q,\omega),
\label{respuesta}
\end{equation}
where $\epsilon=E/m^*_N$ is the initial nucleon energy in units of
$m_N^*$, and $\epsilon_F=E_F/m_N^*$ is the (relativistic) Fermi energy
in the same units.  Moreover we have introduced the energy
distribution of the Fermi gas $n(\epsilon)=
\theta(\epsilon_F-\epsilon)$.  The lower limit, $\epsilon_0$ of the
integral in Eq. (\ref{respuesta}) corresponds to the minimum energy
for a initial nucleon that absorbs energy $\omega$ and momentum $q$.
It can be written as (see Appendix C of ref. \cite{Ama20})
\begin{equation}
\epsilon_0={\rm Max}
\left\{ 
       \kappa\sqrt{1+\frac{1}{\tau}}-\lambda, \epsilon_F-2\lambda
\right\},
\end{equation}
where we have
introduced the
dimensionless variables 
\begin{eqnarray}
\lambda  = \omega/2m_N^*, & 
\kappa   =  q/2m_N^*, &
\tau  =  \kappa^2-\lambda^2. 
\end{eqnarray}
Now we define a mean value of the single-nucleon responses \cite{Cas23}
by averaging with  the
energy distribution $n(\epsilon)$
\begin{eqnarray}\label{def_sn}
\overline{U}_K(q,\omega)
 = \frac{\int^{\infty}_{\epsilon_0} d \epsilon \,  n (\epsilon)
U_K(\epsilon,q,\omega)}{\int_{\epsilon_0}^{\infty} d\epsilon \, n (\epsilon)}.
\end{eqnarray}
Using these averaged single-nucleon responses we can rewrite
Eq. (\ref{respuesta}) in the form
 \begin{eqnarray}
R_K^{QE}(q,\omega)
&=& 
 \frac{V}{(2 \pi)^3}
 \frac{2\pi m_N^{*3}}{q}  2\overline{U}_K
\int_{\epsilon_0}^{\infty}d\epsilon\, n(\epsilon).
\label{respuesta2}
\end{eqnarray}
The superscaling function is defined as
\begin{equation} \label{scaling}
\frac{4}{3}\xi_F f^*(\psi^*)=\int_{\epsilon_0}^{\infty} n (\epsilon) d\epsilon,
\end{equation}
where $\xi_F = \epsilon_F-1 \ll 1$ is the kinetic Fermi energy in
units of $m_N^*$.  Note that this integral only depends on the
variable $\epsilon_0$, which in turn depends on $(q, \omega)$. The
definition (\ref{scaling}) is, except for a factor, similar to that of the
$y$-scaling function $f(y)$ \cite{Sar93,Wes75}, where the scaling
variable $y$ was the minimum moment of the initial nucleon. In this
paper we use the $\psi^*$-scaling variable.  The minimum energy of
the nucleon, $\epsilon_0$, is transformed by a change of
variable into the scaling variable, $\psi^*$, defined as
\begin{equation}
\psi^* = \sqrt{\frac{\epsilon_0-1}{\epsilon_F-1}} {\rm sgn} (\lambda-\tau).
\end{equation}
In relativistic Fermi gas (RFG) and in nuclear matter with RMF 
$1 \le \epsilon_0\le \epsilon_F$ 
and consequently the RFG superscaling
function is zero outside this interval, that corresponds
to $1 < |\psi^*|$ for all nuclei.
 However, in a real nucleus
the momentum is not limited by $k_F$, since nucleons
can have higher momentum, especially correlated nucleons can greatly
exceed the Fermi momentum. This has the effect that the
phenomenological superscaling function is not zero for $|\psi^*|>1$.
Equivalently, the high-energy tail of the phenomenological 
scaling function, for $\psi>1$, can be described as a consequence of
two-particle emission induced by the one-body current. This means that
the observed asymmetry in the scaling function 
can be attributed to the contribution of two-particle
emission processes.

Using $V/(2\pi)^3= N/(\frac83 \pi k_F^3)$ we can write
\begin{equation}
R^{QE}_K = 
\frac{\xi_F}{m_N^* \eta_F^3 \kappa} (Z \overline{U}^p_K+N \overline{U}^n_K)
f^*(\psi^*),
\label{susam} 
\end{equation}
where we have added the contribution of $Z$ protons and $N$ neutrons
to the response functions, and $\eta_F=k_F/m_N^*$.
The SuSAM* approach,
extends the formula (\ref{susam}) 
using a phenomenological scaling function,
obtained from experimental data of $(e,e')$. 
In the SuSAM* of ref. \cite{Mar21}  the 2p2h contribution of MEC
was first subtracted from the inclusive cross section
and then divided by the contribution of the single nucleon. 
\begin{equation}
f_{QE}^* =\frac{ (\frac{d\sigma}{d\Omega d\omega})_{exp}
  -(\frac{d\sigma}{d\Omega d\omega})_{MEC}}{\sigma_M ( v_L r_L + v_T
  r_T) },
\label{fexp}
\end{equation}
where 
\begin{equation} \label{rsn}
r_K = \frac{\xi_F}{m_N^* \eta_F^3 \kappa} 
(Z \overline{U}^p_K+N \overline{U}^n_K).
\end{equation}
Second a selection of the QE data points was done by noting that
approximately half of the data collapse into a point cloud around the
RFG scaling function. This point cloud constitutes the data that can
be considered approximately QE and we reject the rest, which
contribute to inelastic processes. Examples of the selected QE data
are shown in Figs. 1 and 2.  In these figures, it is evident that the
scaled quasielastic data cluster together, forming an asymmetric
thick band that exhibits a tail for high energies. In the results
section, we will show that this tail can be parametrized with a
function that is proportional to the 2p2h phase space.

\section{2p2h response functions with correlated pairs}

Before introducing our parameterization of the 2p2h response due to
the OB current, we first investigate the expected structure of the
2p-2h response functions in a simple model. In this model, the OB
current induces two-particle emission when acting on the 
wave function of a correlated
pair of nucleons.  This is expected to be one of the
contributions to the 2p2h response, although not the only one. In this
section, other effects, such as interference with the two-body
current, are not considered.

Our theoretical motivation is based on the 
RFG and the independent pair approximation or
Bethe-Goldstone equation for the wave function of a pair of
correlated nucleons in presence of the nuclear medium.  Our aim is
not to develop a detailed and exhaustive microscopic model, but rather
to provide some basic outlines in a schematic model, as a
starting point for proposing a formula that accounts for the
fundamental ingredients that determine the 2p2h response.

 For convenience we use here square
brackets in the notation of normalized states within the volume box
$V$, indicating that the states are normalized to unity within the
box
\begin{eqnarray}
|[\np]\rangle \equiv
\frac{e^{i\, \np\cdot\nr}}{\sqrt{V}}\textstyle
|\frac12 s\rangle  \otimes |\frac12 t\rangle ,  
\end{eqnarray}
 while the absence of brackets indicates that the states are
normalized over the entire space to a Dirac's delta of momentum, i.e.
\begin{eqnarray}
|\np\rangle \equiv
\frac{e^{i\, \np\cdot\nr}}{(2\pi)^{3/2}}
\textstyle |\frac12 s\rangle \otimes  |\frac12 t\rangle   
\end{eqnarray}
The previous states carry implicit spin and isospin indices, which we
do not write for convenience and to keep the equations short in order
to enhance clarity. 

In the absence of correlations, the only mechanism of interaction that
can excite a 2p2h state is a two-body current, particularly meson
exchange currents, whose matrix elements in the RFG  
can be written as:
\begin{eqnarray}
\langle [\np'_1\np'_2] | J^\mu(\nq,\omega) | [\nh_1\nh_2] \rangle &=&
\frac{(2\pi)^3}{V^2}
\label{e21}\\
&& \kern -3cm
\delta(\np'_1+\np'_2-\nq-\nh_1-\nh_2)
j^{\mu}(\np'_1,\np'_2,\nh_1,\nh_2).
\nonumber
\end{eqnarray}
where $p'_i>k_F$ and $h_i<k_F$. 
The two-body current functions
$j^{\mu}(\np'_1,\np'_2,\nh_1,\nh_2)$ also depends implicitly 
on spin-isospin indices.

The corresponding inclusive hadronic tensor in the 2p2h channel can be written
as
 \begin{eqnarray}
W^{\mu\nu}_{2p2h}(q,\omega)
&=& 
 \frac{V}{(2 \pi)^9}
\int
d^3p'_1
d^3p'_2
d^3h_1
d^3h_2
\frac{(m^*_N)^4}{E_1E_2E'_1E'_2}
\nonumber\\
&& \times w^{\mu\nu}(\np'_1,\np'_2,\nh_1,\nh_2) 
\nonumber\\
&&
\kern -1cm
\times
\theta(p'_1-k^N_F)
\theta(k^N_F-h_1)
\theta(p'_2-k^{N'}_F)
\theta(k^{N'}_F-h_2)
\nonumber\\
&& \times \delta(E'_1+E'_2-E_1-E_2-\omega)
\nonumber\\
&&
\times
\delta(\np'_1+\np'_2-\nq-\nh_1-\nh_2), 
\end{eqnarray}
where $E_i$,  $E'_i$ are the on-shell energies of nucleons 
with momenta $\nh_i$, $\np'_1$, and 
with relativistic 
effective mass $m_N^*$,
The function $w^{\mu\nu}(\np'_1,\np'_2,\nh_1,\nh_2)$ represents the
hadron tensor for a single 2p2h transition,
summed up over spin and isospin,
\begin{eqnarray}
\kern -8mm
w^{\mu\nu}(\np'_1,\np'_2,\nh_1,\nh_2) &=& 
\nonumber\\
&& \kern -3cm
=\frac{1}{4}
\sum_{s_1s_2s'_1s'_2}
\sum_{t_1t_2t'_1t'_2}
j^{\mu}(1',2',1,2)^*_A
j^{\nu}(1',2',1,2)_A \, . 
\label{elementary}
\end{eqnarray}
where the two-body current
function is antisymetrized  
\begin{equation} \label{anti}
j^{\mu}(1',2',1,2)_A
\equiv j^{\mu}(1',2',1,2)-
j^{\mu}(1',2',2,1) \,.
\end{equation}
 The factor $1/4$ in Eq.~(\ref{elementary}) accounts for the
 anti-symmetry of the two-body wave function with respect to exchange
 of momenta, spin and isospin quantum numbers, to avoid double
 counting of the final 2p2h states.
 
Let's now consider the matrix element of the one-body current
$J^\mu(\nq,\omega) = J^\mu_{OB} = \sum_i J^\mu_i$, where the $J^\mu_i$
acts on the i-th particle. We have 
\begin{equation}
\langle \np'_1\np'_2 | J^\mu_{OB} | \nh_1\nh_2 \rangle=
\langle \np'_1\np'_2 | J^\mu_1 + J^\mu_2  | \nh_1\nh_2 \rangle =0
\end{equation}
 Indeed, a one-body current cannot produce
2p2h excitations due to the orthogonality of plane waves
$\langle\np'_i|\nh_j\rangle=0$. 

The situation changes when we consider that the two states
$|\nh_i\rangle$ are interacting in the medium, with a short-range NN
potential $V_{NN}$ that produces scattering to unoccupied states. If we
turn-on the NN interaction the wave function $|\nh_1\nh_2\rangle$ of
the pair is modified, but due to Pauli blocking it can only acquire
momentum components above the Fermi momentum. 
In the independent pair approximation 
 the wave function $|\nh_1\nh_2\rangle$
is replaced by the correlated wave function $|\Phi_{\nh_1\nh_2}\rangle$
\begin{eqnarray}
 |\nh_1,\nh_2\rangle \rightarrow
  |\Phi_{\nh_1\nh_2}\rangle = |\nh_1,\nh_2\rangle +   
|\Delta\Phi_{\nh_1\nh_2}\rangle
\end{eqnarray}
where $ |\Delta\Phi_{\nh_1\nh_2}\rangle$ only has high momentum
components.
In the independent pair approximation the wave function is obtained
from the solution of the BG equation for the correlated
wave function \cite{Wal95}.
Since the 
NN interaction conserves the total momentum, 
the correlated part of the wave function in momentum space verifies
\begin{equation}  \label{deltapsi}
\langle \np_1\np_2|\Delta\Phi_{\nh_1\nh_2}\rangle
=
\delta(\np_1+\np_2-\nh_1-\nh_2)
\Delta\varphi_{\nh_1\nh_2}(\np)
\end{equation}
where 
$\np=\frac12(\np_1-\np_2)$ is the relative momentum of the nucleon pair, and $\Delta\varphi_{\nh_1\nh_2}(\np)$
 is the relative wave function
of $\Delta\Phi_{\nh_1\nh_2}$. From the BG equation, 
it is given  by \cite{Rui17b}
\begin{equation}\label{relativa}
\Delta\varphi_{\nh_1\nh_2}(\np)
=
 \frac{\theta(p_1-k_F)\theta(p_2-k_F)}{h^2-p^2}\,
\langle \np| 2\mu V_{NN} | \varphi_{\nh_1\nh_2}\rangle
 \end{equation}
where 
$\varphi_{\nh_1\nh_2}(\np)$
 is the relative wave function of  $\Phi_{\nh_1\nh_2}$,
 $\mu=\frac{m_N}{2}$ is the reduced mass of the two-nucleon
system and $\nh=(\nh_1-\nh_2)/2$ the initial relative momentum, while 
 $\np_1=(\nh_1+\nh_2)/2 +\np$ and $\np_2=(\nh_1+\nh_2)/2 -\np$.
The Pauli blocking functions in Eq. (\ref{relativa}) ensure
that the wave function has high-momentum components.  As we can see,
it is not possible to remove  the dependence on the
total momentum $\nh_1+\nh_2$ appearing in the Pauli-blocking step
functions. Therefore a dependence of short-range correlations on the
CM momentum of the pair appears.

Now, by applying a one-body current to a system of two correlated
nucleons, it is possible to generate a two-particle state that exists
above the Fermi level.  The 2p-2h matrix element of the one-body
current is computed in Appendix \ref{apendiceA}.  It can be written 
similarly to Eq. (21) 
\begin{eqnarray} \label{matrix}
\langle [\np'_1\np'_2] | J^\mu_{OB}(\nq) | [\Phi_{\nh_1\nh_2}] \rangle &=&
\frac{(2\pi)^3}{V^2}
\nonumber\\
&& \kern -4cm
\delta(\np'_1+\np'_2-\nq-\nh_1-\nh_2)
j^{\mu}_{cor}(\np'_1,\np'_2,\nh_1,\nh_2).
\end{eqnarray}
It is noteworthy that the matrix element describing the effect of
the one-body  current on the wave function of two correlated nucleons is
formally similar to the matrix element of a two-body correlation
current, which is represented by the function $j^\mu_{cor}$
\begin{eqnarray}
j^{\mu}_{cor}(\np'_1,\np'_2,\nh_1,\nh_2)
&=& 
\nonumber\\
&&
\kern -3cm
(2\pi)^3
j^{\mu}_{OB}(\np'_1,\np'_1-\nq)
\Delta\varphi_{\nh_1\nh_2}\left(\np'-\frac{\nq}{2}\right)
\nonumber\\
&& 
\kern -3cm
\mbox{}+
(2\pi)^3
j^{\mu}_{OB}(\np'_2,\np'_2-\nq)
\Delta\varphi_{\nh_1\nh_2}\left(\np'+\frac{\nq}{2}\right)
\label{jsimple}
\end{eqnarray}
where $\np'=(\np'_1-\np'_2)/2$ is the relative momentum of the final particles.

\begin{figure}
\includegraphics[scale=0.8,bb=166 519 444 694]{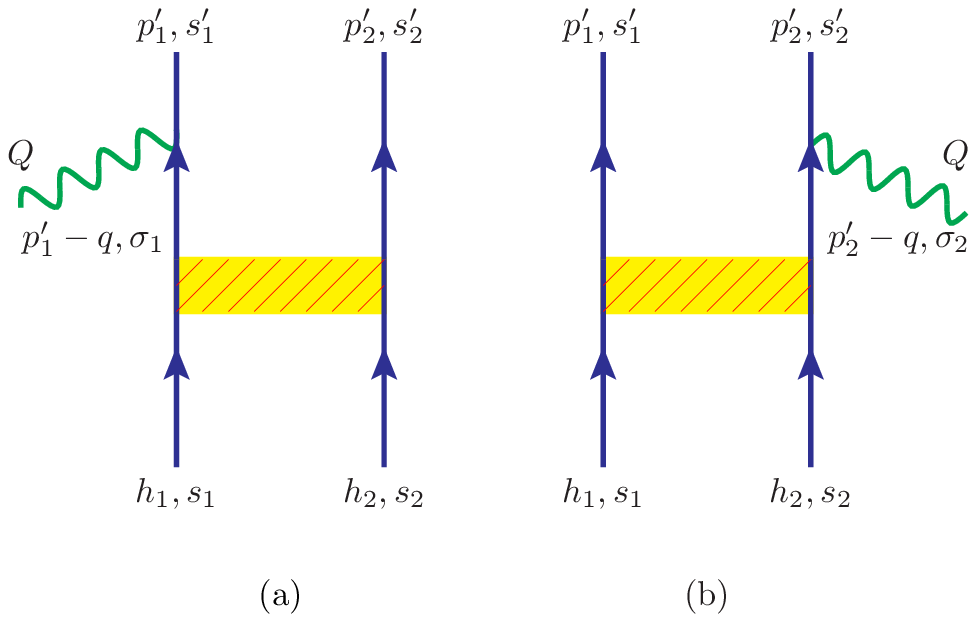}
  \caption{ 
Diagram illustrating the correlation current, Eq
    (\ref{jsimple}). The spins of the final particles are $s'_1$ and $s'_2$,
whereas $\sigma_1,\sigma_2$ denotes one of the spin components of
    the relative wave function of high momentum,
    represented by the shaded rectangle.
}\label{diagcor}
\end{figure}

The correlation current is the sum of the products of the OB current
multiplied by the high-momentum wave function of the initial
correlated pair.  This is illustrated in the diagrams of
Fig. \ref{diagcor}. The shaded rectangle represents the
correlations resulting in the production of high-momentum components
  of the correlated pair $(\nh_1,\nh_2)$.
One of the high-momentum nucleons 
absorbs a photon with momentum $q$,
while the other nucleon is emitted through its interaction with the
first nucleon.
In Eq. (\ref{jsimple})  the products are, in fact,
a multiplication of spin matrices, although we have left this out of
equation (30) for clarity. The complete formula dependent on spin is
provided in the appendix \ref{apendiceA}, Eq. (\ref{jcor}).  Note also that
the current depends on isospin. Hence, when the initial state consists
of a proton-neutron pair, the current acting on the first particle
must correspond to the proton current, while the current acting on the
second particle should correspond to the neutron current.

By inserting the correlation current (\ref{jsimple}) in Eq. (23)
the hadronic tensor for a 2p2h
transition is
obtained. The isospin sums in Eq. (23) can be written as sums over  
pp, pn, and nn correlated pairs. Therefore
\begin{equation}
 w^{\mu\nu}(\np'_1,\np'_2,\nh_2,\nh_2)=
w^{\mu\nu}_{pp}+
w^{\mu\nu}_{np}+
w^{\mu\nu}_{nn}.
\end{equation}
Writing explicitly the isospin indices,  $N, N'=p,n$, 
the diagonal components of the hadronic tensor 
$w^{\mu\mu}_{NN'}$ are the following
\begin{eqnarray}
w^{\mu\mu}_{NN'}(\np'_1,\np'_2,\nh_1,\nh_2)
&=& 
\nonumber\\
&&
\kern -3cm
(2\pi)^6
\left|
j^{\mu}_{N}(\np'_1,\np'_1-\nq)
\Delta\varphi^{NN'}_{\nh_1\nh_2}\left(\np'-\frac{\nq}{2}\right)
\right|^2
\nonumber\\
&& 
\kern -3cm
\mbox{}+
(2\pi)^6
\left |j^{\mu}_{N'}(\np'_2,\np'_2-\nq)
\Delta\varphi^{NN'}_{\nh_1\nh_2}\left(\np'+\frac{\nq}{2}\right)
\right|^2
\nonumber\\
&&
\kern -3cm
\mbox{}+
2(2\pi)^6\mbox{Re}\left\{
j^{\mu*}_{N}(\np'_1,\np'_1-\nq)
\Delta\varphi^{NN'*}_{\nh_1\nh_2}\left(\np'-\frac{\nq}{2}\right)
\right.
\nonumber\\
&& 
\kern -3cm
\left.
\mbox{}\times
j^{\mu}_{N'}(\np'_2,\np'_2-\nq)
\Delta\varphi^{NN'}_{\nh_1\nh_2}\left(\np'+\frac{\nq}{2}\right)
\right\}
\label{wsimple}
\end{eqnarray}

In the independent pair approximation the 2p2h hadronic tensor
$W^{\mu\nu}_{NN'}(q,\omega)$ due to SRC, is defined as the integral of
the correlated-pair tensor
$w^{\mu\nu}_{NN'}(\np'_1,\np'_2,\nh_2,\nh_2)$ over the momentum space
of 2p2h excitations, using Eq. (22).  The computation of this tensor
falls outside the scope of the present work. The proposed approach for
such a calculation would require solving the Bethe-Goldstone equation
for every nucleon pair $\nh_1, \nh_2$ while considering the center of
mass of the two particles $\nh_1+\nh_2 \ne 0$, and performing a
seven-dimensional integration.  In reference \cite{Rui17b}, the BG
equation was solved for the particular case of back-to-back nucleons,
$\nh_1+\nh_2=0$, moving along the $z$-axis in a multipole expansion
with a potential $V_{NN}$ fitted to NN scattering data \cite{Nav13}.
Although the computation seems feasible, incorporating the
center of mass for particles moving in any direction requires careful
consideration of various technical details. 

The phenomenological approach of scaling analysis relies on the
factorization approximation of the single-nucleon response. In the
subsequent section, we will explore the consequences of adopting this
approximation in the model introduced in the current section. We will
further generalize it to derive an empirical factorized formula that
can be applied within the scaling approach.  However, it is important
to note that the factorization approximation is a simplification and
may overlook certain many-body effects that can play
a role in the response. Therefore, the empirical factorized
formula should be seen as an approximation that captures the main
trends observed in the scaling behavior rather than a complete
microscopic description.

\section{Semiempirical 2p2h response functions}
The 2p2h response functions for two-body operators (MEC) was explored
in Ref. \cite{Mar21b}.  Our analysis revealed that the responses of a
pair can be factorized out of the integral (22) with reasonable
approximation. By doing so, we derived a semi-empirical expression for
the MEC responses, with the coefficients to be determined by later
fitting. 
In this work, we make the assumption that a similar factorization
occurs for the 2p2h response induced by the one-body (OB)
current. This implies that the 2p2h response can be proportional to
the phase-space integral of two nucleons.
  In the framework of the RMF of nuclear matter
the 2p2h phase-space function is given by the integral
 \begin{eqnarray}
F_{NN'}(q,\omega)
&=& 
\int
d^3p'_1
d^3p'_2
d^3h_1
d^3h_2
\frac{(m^*_N)^4}{E_1E_2E'_1E'_2}
\nonumber\\
&&
\kern -1cm
\times
\theta(p'_1-k^N_F)
\theta(k^N_F-h_1)
\theta(p'_2-k^{N'}_F)
\theta(k^{N'}_F-h_2)
\nonumber\\
&& \times \delta(E'_1+E'_2-E_1-E_2-\omega)
\nonumber\\
&&
\times
\delta(\np'_1+\np'_2-\nq-\nh_1-\nh_2),  \label{ps}
\end{eqnarray}
where $N,\ N'$ can be proton or neutron, depending on the initial
correlated state pp, pn or nn
and $k_F^N$ is the Fermi momentum of the nucleon of  $N$ kind.
The phase space function is roughly proportional to the
 number of 2p2h excitations allowed by
the kinematic for momentum and energy transfer $(q,\omega)$. 

The concept of factorization in the 2p2h channel bears some
resemblance to that employed in the 1p1h channel, where the response
was written as a single-nucleon averaged response multiplied by the
superscaling function. 
In the 2p2h case, we will also  define the
two-nucleon tensor averaged over the momentum space of 2p2h excitations,
by dividing the hadronic tensor over the phase space
function as follows
\begin{eqnarray}
\overline{w}^{\mu\nu}_{NN'}(q,\omega)\equiv 
\frac{W^{\mu\nu}_{NN'}(q,\omega)}{ \frac{V}{(2 \pi)^9}F_{NN'}(q,\omega)}
\end{eqnarray}
In the definition given in Eq. (34) $W^{\mu\nu}_{NN'}(q,\omega)$ 
can be in general the exact
hadronic tensor for 2p2h emission including the one-body current
and possibly other interference contributions, not only the correlations. 
This leads to an exact factorization of the 2p2h hadronic tensor
as the averaged two-nucleon tensor multiplied by the phase
space.
\begin{eqnarray}
W^{\mu\nu}_{NN'}(q,\omega)=
 \frac{V}{(2 \pi)^9}F_{NN'}(q,\omega)\overline{w}^{\mu\nu}_{NN'}(q,\omega)
\end{eqnarray}
In this formula the averaged two-nucleon tensor accounts for the
correlations and other interaction contributions such as interferences
with the two-body current, while the phase space does not contain any
information about them.  The phase space is exclusively related to the
kinematics of independent particles without interactions in the Fermi
gas.

To proceed further, we analyze the specific case of the hadronic
tensor in the model of independent pairs that was introduced in the
previous section.  The average of the tensor $w^{\mu\nu}_{NN'}$ in
equation (\ref{wsimple}) is typically complicated.  Two approximations
lead to the simplification of this term.  We will make the assumption
that the average of the product can be written as the product of
averages, and neglect on the average the interference term between the
two particles, given by the last term in Eq. (\ref{wsimple}). In this
way we obtain for the diagonal components
\begin{eqnarray}
\overline{w}^{\mu\mu}_{NN'}(q,\omega)
&\simeq& 
(2\pi)^6
\overline{
\left(\left|j^{\mu}_{N}\right|^2+\left|j^{\mu}_{N'}\right|^2
\right)}
\times
\overline{\left|
\Delta\varphi^{NN'}_{\nh_1\nh_2}\right|^2},
\nonumber\\
\label{esquema}
\end{eqnarray}
where we use the notation
\begin{equation}
\overline{\left|j^{\mu}_{N}\right|^2}
\equiv
 \overline{\left|j^{\mu}_{N}(\np'_1,\np'_1-\nq)\right|^2}
=
 \overline{\left|j^{\mu}_{N}(\np'_2,\np'_2-\nq)\right|^2}
\end{equation}
\begin{eqnarray}
\overline{\left|\Delta\varphi^{NN'}_{\nh_1\nh_2}\right|^2}
&\equiv&
\overline{
\left|\Delta\varphi^{NN'}_{\nh_1\nh_2}\left(\np'+\frac{\nq}{2}\right)
\right|^2}
\nonumber\\
&=&
\overline{
\left|\Delta\varphi^{NN'}_{\nh_1\nh_2}\left(\np'-\frac{\nq}{2}\right)
\right|^2}.
\end{eqnarray}
Therefore, by performing the factorization approximation of the OB
current in the independent-pair model and neglecting the
interferences, we find that the corresponding hadronic tensor, based
on Eqs (35,36), includes the product of three factors: the 2p2h
phase-space integral, the averaged response of a single nucleon
(coming from the one-body current), and a factor related to the
averaged high-momentum distribution of a pair. In the semiempirical
formula for the 2p2h response we assume the same factorization, we
will replace this factor with an adjustable parameter that is not
necessarily associated solely with correlations, as it will include
contributions and interferences with other processes, particularly
MEC. Hence, we propose the following semi-empirical formula for the
2p2h responses induced by the OB current in electron scattering.
\begin{eqnarray}
R^K_{\rm 2p2h} 
&=&
 \frac{V}{(2 \pi)^9}
\frac{1}{m_N^{2}  m_\pi^{4}}\frac{2}{A(A-1)}\times
\nonumber\\
&& \left[
 \frac{Z(Z-1)}{2} F_{pp}(q,\omega)  c_K^{pp}(q) (2\overline{U}_K^p) \right.
\nonumber\\
&+& NZ F_{pn}(q,\omega)  c_K^{pn}(q)  (\overline{U}_K^p + \overline{U}_K^n ) 
\nonumber\\
&+& \left. 
\frac{N(N-1)}{2} F_{nn}(q,\omega)  c_K^{nn}(q)  (2\overline{U}_K^n) 
\right].
\label{asimetrica}
\end{eqnarray} 
The mass of the pion $(m_\pi)^4$ in the denominator has been
introduced for convenience by analogy to MEC responses
\cite{Mar21b} and  
the mass of the nucleon $m_N^2$ in the denominator 
is set so that the 2p2h parameters $c_K^{NN'}(q)$
are dimensionless.

In Eq. (\ref{asimetrica}), we have separated the contributions of the
pairs, pp, pn and nn.  Each of them is proportional to the sum of the
corresponding single-nucleon responses, $\overline{U}^p_K$ and/or
$\overline{U}^n_K$.  This is because the photon can be absorbed by
both nucleons of the pair.  Also, each contribution has been
multiplied by the number of pairs pp, pn or nn and divided by the
total number of pairs, $A(A-1)/2$, to take into account asymmetric
matter $Z\ne N$.  This approach follows the lines of the model of
ref. \cite{Ngu20}.

Since we expect that the 2p2h parameters $c_K(q)$ are related, at
least partially, to the high momentum components of a nucleon pair,
they should strongly depend on isospin, given that the high momentum
components of proton-neutron dominate over proton-proton and
neutron-neutron.  Here we will assume that they are proportional
$c_K^{pp}(q) = c_K^{nn}(q) = \alpha c_K^{pn}(q)$, where $\alpha$ is a
small constant.  Experiments on $^{12}$C have reported a number of np
pairs $18$ times larger than their pp counterparts \cite{Jef07,Sub08}.
Then a reasonable value is $\alpha=1/18$.

Is also reasonable to assume that the $c_K^{NN'}$ coefficients are the same for
both the longitudinal and transverse response, 
$c_L^{NN'}=c_T^{NN'}=c^{NN'}$. 
This assumption greatly
simplifies the semiempirical formula  
because it only depends on two parameters $c^{pn}$ and $\alpha$.
It is
important to note that this assumption may not hold true in all cases,
and care should be taken when interpreting results obtained using this
simplification.

In the particular case of the factorized independent-pair model 
neglecting the interferences, 
based on the equation (36), the 2p2h parameter $c^{pn}(q)$ 
is related to the average momentum distribution of a proton-neutron pair
 in the 2p2h excitation
\begin{equation}
\frac{c^{pn}(q)}{m_N^2m_\pi^4}
\simeq (2\pi)^6\overline{
\left(\sum_{s_1s_2s'_1s'_2}
\left|\Delta\varphi_{\nh_1\nh_2}^{pn}(\np'+\nq/2)\right|^2\right)
}.
\end{equation}
However, in the real case, this may not necessarily hold true because
there are other mechanisms to consider, such as interference with
two-body currents, FSI, etc. Therefore, in
the present phenomenological approach, the 2p2h parameters are
considered as adjustable quantities to reproduce the tail of the
scaling function. In other words, they quantify to some extent the
asymmetry of the phenomenological scaling function (see next section).

In the case of symmetric nuclei, $N=Z$, Eq. (\ref{asimetrica}) reduces
to
\begin{equation}
R^K_{\rm 2p2h} 
=
 \frac{V}{(2 \pi)^9}
 F(q,\omega) \frac{Z+\alpha(Z-1)}{2Z-1}
 \frac{c^{pn}(q)}{m_N^{2}  m_\pi^{4}}
  (\overline{U}_K^p + \overline{U}_K^n ).
\end{equation} 
From the response functions we can write
the semiempirical formula for the inclusive
cross section in the 2p2h channel induced by the one-body current 
\begin{eqnarray}
\left( \frac{d \sigma}{d \Omega' d\epsilon'} \right)_{\rm 2p2h}^{em} 
&=&
  \frac{\sigma_{\rm Mott} V}{(2 \pi)^9}
\frac{F(q,\omega)c^{pn}(q)}{m_N^{2}  m_\pi^{4}}
\frac{Z+\alpha(Z-1)}{2Z-1}
\nonumber\\
&&
\kern -1cm \left [  v_L (\overline{U}_L^p + \overline{U}_L^n )
+ v_T (\overline{U}_T^p + \overline{U}_T^n)\right].
\label{semi1}
\end{eqnarray} 
In the next section we use the $^{12}{\rm C}(e,e')$ cross section data
to fit the 2p2h parameter $c^{pn}(q)$ for each $q$.

In the case of CC neutrino scattering the semiempirical formula
extends naturally by replacing the electromagnetic single nucleon
responses with the corresponding to the $n(\nu_\mu,\mu)p$ or
$p(\overline{\nu}_\mu,\mu^+)n$, assuming the same relation between the 
2p2h parameters of the $nn$ and ($pp$ in the case of anti-neutrino) 
 pairs to the $np$ pairs in the initial state
$c^{nn}=c^{pp}= \alpha c^{pn}$ 
\begin{eqnarray}
\left( \frac{d \sigma}{d \Omega' d\epsilon'} \right)_{\rm 2p2h}^{\nu} 
&=&
\frac{\sigma_0 V}{(2 \pi)^9}
\frac{ F(q,\omega)  c^{pn}(q)}{m_N^{2}  m_\pi^{4}}
\frac{Z+\alpha(Z-1)}{2Z-1}
\nonumber\\
&&\left[ V_{CC} \overline{U}_{CC}
+2 V_{CL} \overline{U}_{CL}
+ V_{LL} \overline{U}_{LL} \right.
\nonumber\\
&&+ \left.
 V_{T} \overline{U}_{T}
\pm 2 V_{T'} \overline{U}_{T'}\right],
\label{semi2}
\end{eqnarray} 
where $\sigma_0$ is  
\begin{equation}
\sigma_0= 
\frac{G^2 \cos^2 \theta_c}{4\pi^2}\frac{k'}{\epsilon}
[(\epsilon+\epsilon')^2-q^2],
\end{equation}
with $G$ the Fermi constant and $\theta_c$ the Cabibbo angle. The
lepton coefficients for neutrino scattering $V_K$ are given in
Ref. \cite{Ama20}. The average single nucleon responses 
$\overline{U}_K$ are given in Appendix B.
Here we use a new version of these responses using the definition (12)
with the distribution $n(\epsilon)$ obtained from the
phenomenological scaling function. This differs from the traditional 
definition \cite{Rui18} only for $\psi>1$, where the contribution of
the nucleons with momentum greater than $k_F$ is correctly taken into
account, while the traditional definition is an extrapolation of the
RFG where the nucleons are limited by $k_F$ \cite{Cas23}.

\begin{figure}
\includegraphics[scale=0.55,bb=80 285 480 770]{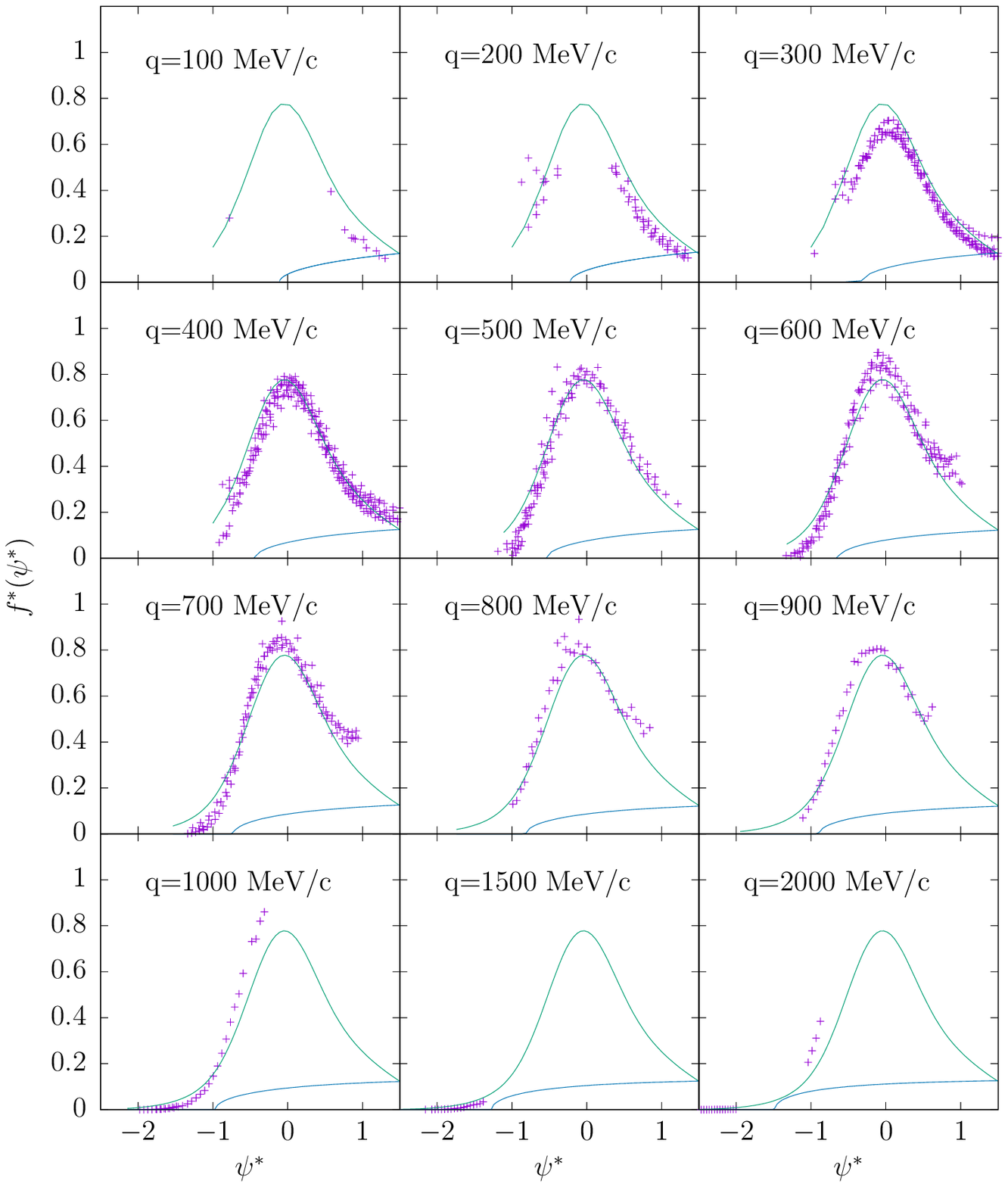}
  \caption{ Scaling function of the SuSAM* model (green) compared to
    the contribution $f_{2p2h}^*$ in the 2p2h channel (blue) for
    several values of $q$, as a function of the scaling variable
    $\psi^*$.  For each $q$ we show only the scaled $^{12}$C data for
    $q\pm 50$ MeV/c. The original data are taken from
    \cite{archive,archive2} }
\end{figure}

\begin{figure}

  \centering
    \includegraphics[width=8cm,bb=70 500 500 800]{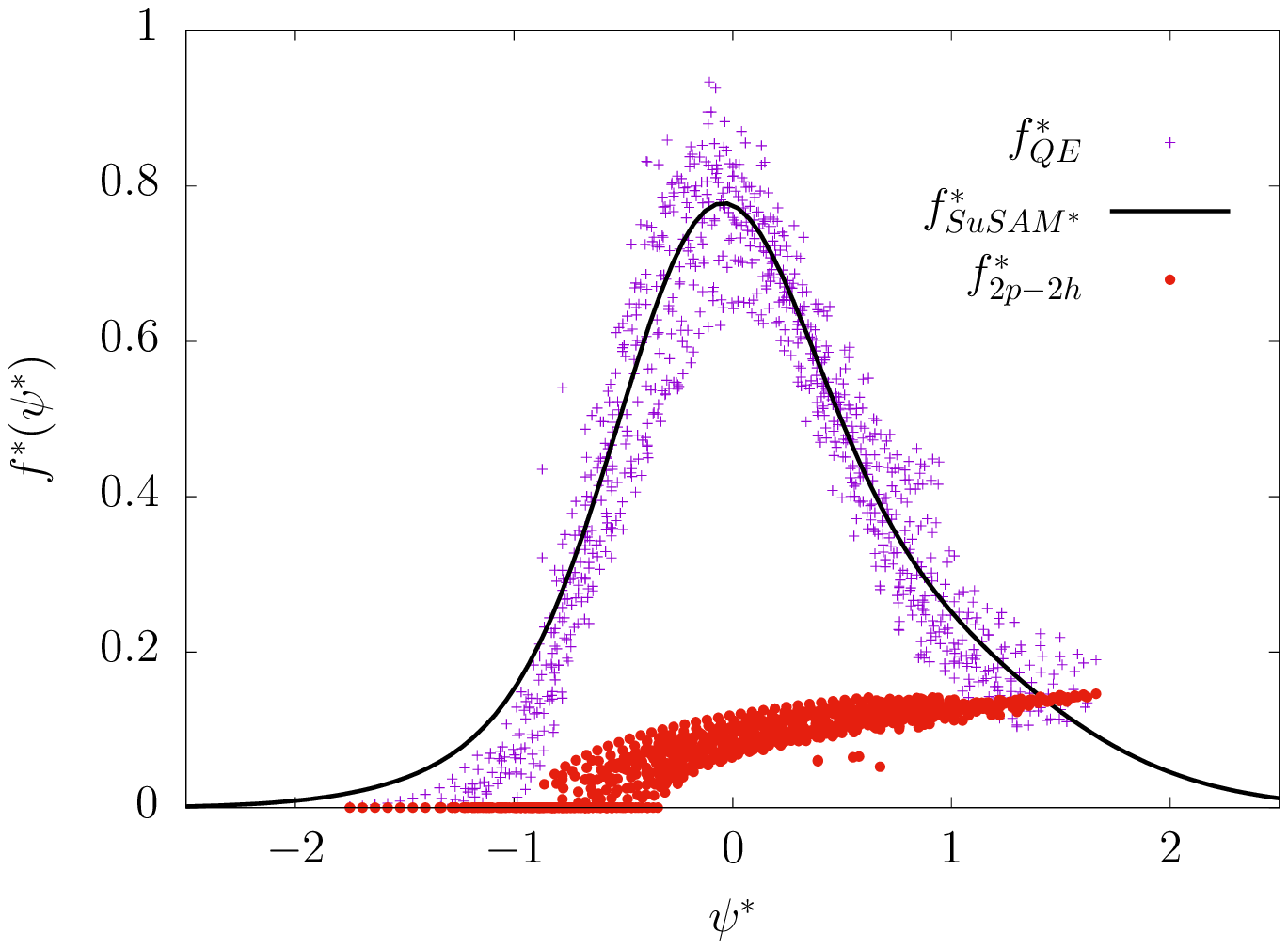}
  \caption{
Experimental data of the scaling function $f^*_{QE}$ 
compared to the 2p2h contribution.
Only data with $q\leq 1000$ MeV/c are shown. The solid line is the SuSAM* fit.
 }
  \label{fig0}
\end{figure}

Finally, in the semi-empirical formula will make use of the 
following approximation for the phase-space function  \cite{Sim14b}:
\begin{equation} 
  F_{NN'}(q,\omega)=
\left(4\pi k_F^N k_F^{N'}\right)^3
\frac{m_N^{* 2}}{18} \sqrt{1-\frac{4m_N^{*2}}{(2m_N^*+\omega )^2-q^2}}.
\label{analytical}
\end{equation}
Eq. (\ref{analytical}) makes use of the frozen nucleon approximation 
to compute the integral (\ref{ps}).
The exact phase space was studied in depth in ref \cite{Sim14} as a
function of $(q,\omega)$ for the kinematics of interest for neutrino
experiments, $q \sim 1$ GeV, around the quasielastic peak.
For these kinematics it was seen that the
frozen approximation gives results very close to the exact value. The
frozen approximation was also found to be quite accurate in the 2p2h
MEC responses \cite{Sim17b}. 
and in the semi-empirical
formula of the MEC responses \cite{Mar21b}.

\section{Results}

In this section, we extend the superscaling model introduced in
Section II and perform a new analysis of the (e,e') data assuming that
the scaling function contains a contribution from 2p2h that produces
the tail of the quasielastic scaling data. Therefore, in the extended
scaling model, we replace the quasielastic superscaling function
$f^*_{QE}$ with $f^*_{1p1h}+f^*_{2p2h}$. The 2p2h component of the
scaling function is parameterized using the semi-empirical formula for
the 2p2h response discussed in the previous section, and thus it is
proportional to the phase space function $F(q,\omega)$. This can be
seen as an alternative to the traditional superscaling analysis, which
typically parametrizes the 2p2h component using Gaussian functions,
while in this approach, we impose an additional condition.

The extended scaling analysis is carried out in three steps: i) A
preliminary scaling analysis is performed as usual to obtain the
function $f^*_{QE}$, 
after subtracting the contribution of the MEC from the data, Eq. (17). 
 ii) For each value of $q$, the 2p2h parameter is
adjusted such that the function $f^*_{2p2h}$ reproduces the high-energy
tail of the scaling function $f^*_{QE}$.  iii) The contribution of
$f^*_{2p2h}$ is subtracted from the data, and a new scaling analysis is
performed to extract the 1p1h scaling function $f^*_{1p1h}$.

After this procedure, it is observed that the resulting $f^*_{1p1h}$
function is compatible with a symmetric function, indicating that the
assumed dependence for the tail is appropriate. In other words, the tail
of the scaling function is consistent with the phase space function,
and this is in agreement with the hypothesis that the
tail is generated by 2p2h excitations. This is why the symmetric
scaling function is referred to as $f_{1p1h}$ because it no longer
includes 2p2h contributions, which have been subtracted from the
analysis.

For each value of $q$, the coefficient $c^{pn}(q)$ is fitted from
the phenomenological scaling function $f^*_{QE}$ under the hypothesis
that the high energy tail ($\psi^*>1$) is produced 
mainly by 2p2h excitations,
In Fig. 2 we show the
experimental data of $f^*_{QE}$ for 
$^{12}$C and for different $q$-values. 
These data have been obtained from the inclusive cross
section by subtracting the 2p2h MEC contribution and dividing by the
single nucleon cross section, Eq. (\ref{rsn}).  The Fermi momentum is
$k_F=225$ MeV/c and $M^*=m^*_N/m_N=0.8$ \cite{Mar21}. 
The errors of these parameters were estimated in ref \cite{Ama18}
in a $\chi^2$ fit. They are $\Delta k_F= 8$ MeV/c and $\Delta M^*$=0.044.
 From the figure we see
that the scaling is only approximate and the data are concentrated in
a narrow band. 
The band is shown in Fig. 3 for $q\leq 1000$ MeV/c. 
The scaling model is based on two assumptions; first the
single-nucleon factorization, which can be done assuming that the
FSI are small in the quasielastic
region. The second assumes that the scaling function depends solely on
the scaling variable, which is a consequence of approximating the
nuclear system by an infinite Fermi gas.  The small scaling violation
can be attributed to nuclear effects that breaks this approximation
such as FSI, finite size and off-shell effects.
Still the band of Fig. 3 is well
defined following the shape of an asymmetric bell, with an evident tail
on the right. 
 The phenomenological scaling function of the SuSAM* model is obtained
by fitting a sum of Gaussians to this band:
\begin{equation} \label{central}
f_{QE}^*(\psi^*) = 
 a_3e^{-(\psi^*-a_1)^2/(2a_2^2)}+ b_3e^{-(\psi^*-b_1)^2/(2b_2^2)}.
\end{equation}
Note that there are no QE data above $\psi^*=1.5$. Then 
it is not possible to know from the data how the tail 
would extend above this value.
This is because the region 
$\psi^* > 1.5$ requires high values of the energy far from the QE region.
For low $q$ there are not enough cross section data in this region, and for 
higher momentum transfer, $q> 500$ MeV/c  we 
enter the inelastic zone with pion emission and the data no longer
scale.  

If we assume that the tail of the scaling function is primarily due to
the emission of two nucleons, then the semiempirical formula offers a
theoretical framework to extrapolate the model to high values of
$\psi^*$ (or high energy transfer $\omega$). This hypothesis is
supported by numerous calculations \cite{Co01,Co98,Ama10,Cuy16} which
have shown that 2p2h inclusive responses exhibit an increasing
behavior with energy transfer, contributing to the formation of a tail
in the cross section at high energies. Although these calculations are
based on various MEC and/or SRC models using different approaches,
they all qualitatively resemble the 2p2h phase space, which aligns
with the semiempirical approach proposed in this work.

\begin{table}
\begin{tabular}{lllllllllll}
\hline\hline 
$q$ [GeV/c]    & 0.1  & 0.2  & 0.3 & 0.4 & 0.5 & 0.6 & 0.7 & 0.8 & 0.9 & 1  
\\ \hline\hline 
$c^{pn}(\alpha=0)$ & 8067 & 2750 & 1558 & 1008 & 733 & 550 & 458 & 367 & 312 & 275  \\
$c^{pn}(\alpha=\frac{1}{18})$ & 7710 & 2628 & 1489 & 964 & 701 & 526 & 438 & 350 & 298 & 263  
\\ \hline\hline
\end{tabular}
\caption{ Coefficients of the semiempirical formula for different
  values of the parameter $\alpha$. The estimated error of $c^{pn}(q)$ is
  38\%.}
\label{table}
\end{table}

\begin{figure}
  \includegraphics[width=8cm,bb=70 500 500 800]{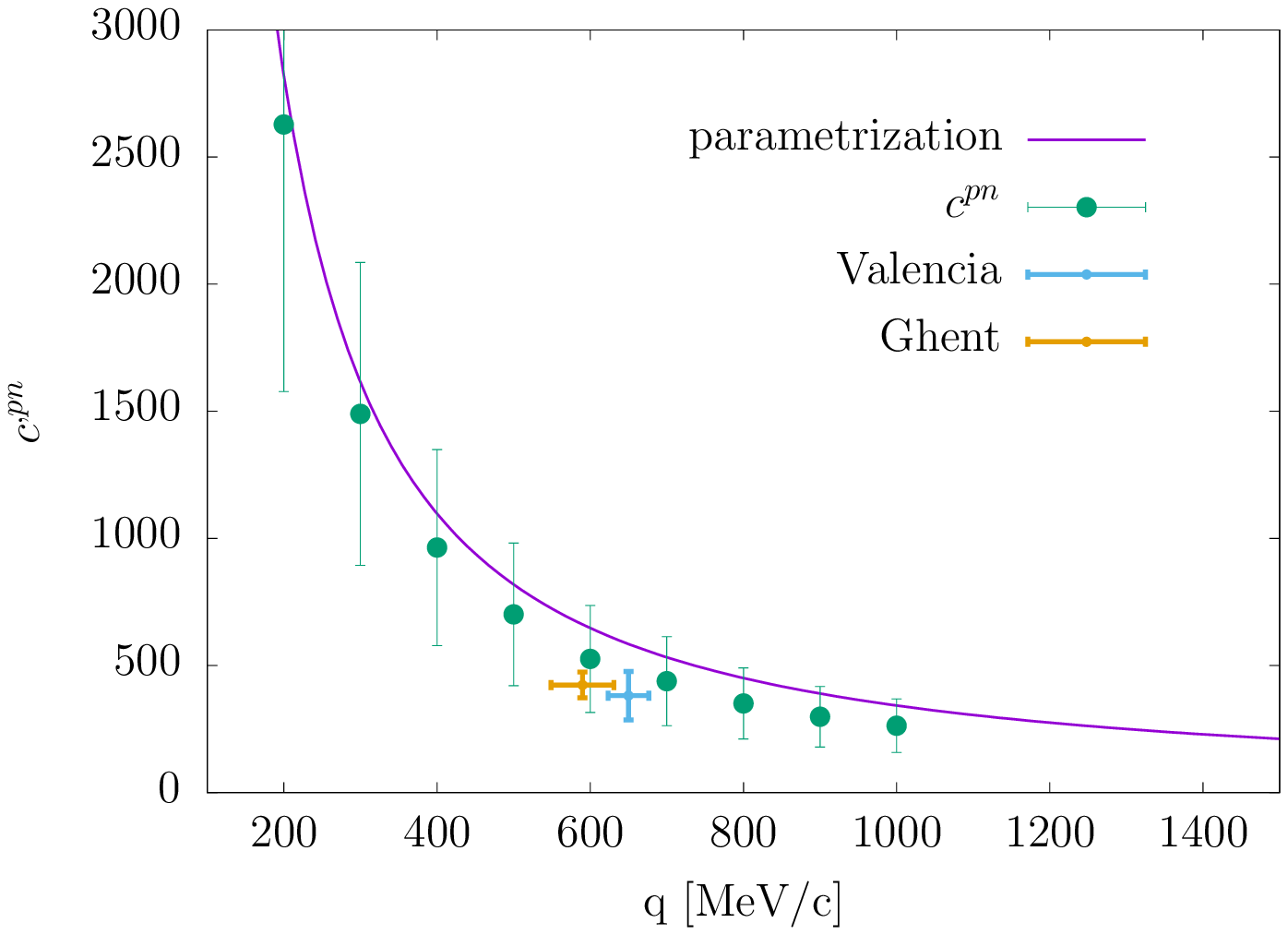}
  \caption{ Fitted values of the coefficients of the semiempirical
    formula with error bars compared to the interpolating function
    of Eq. (\ref{interpol}).  }
\end{figure}

\begin{figure}
    \includegraphics[width=8cm,bb=70 500 500 800]{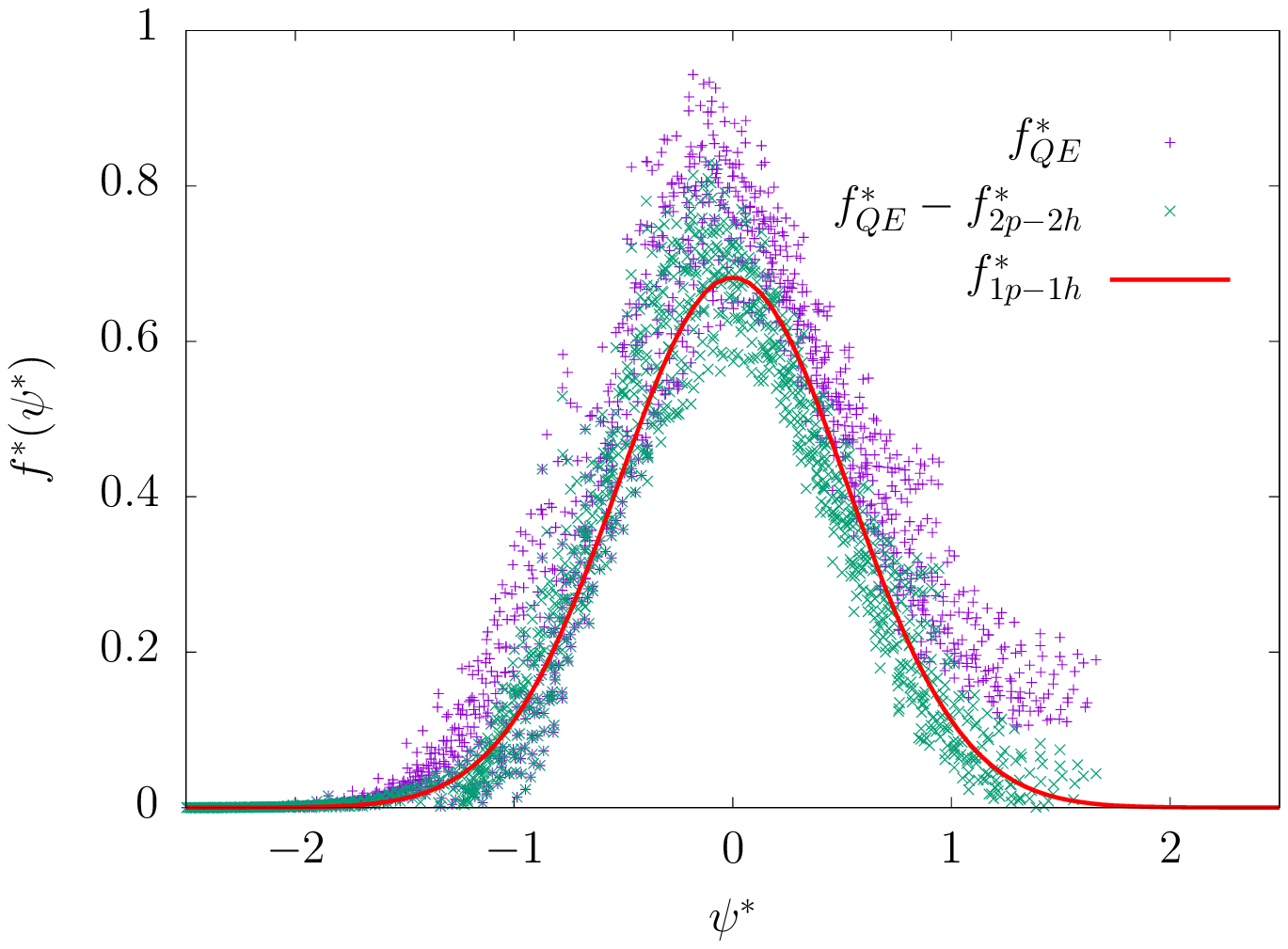}
  \caption{
Experimental values of the 
scaling function of $^{12}$C before and after the 
subtraction of the 2p2h contribution.
The solid line is the new SuSAM* 1p1h scaling function.
}
\end{figure}

\begin{figure*}
  \centering
\includegraphics[scale=0.6,bb=50 380 510 770]{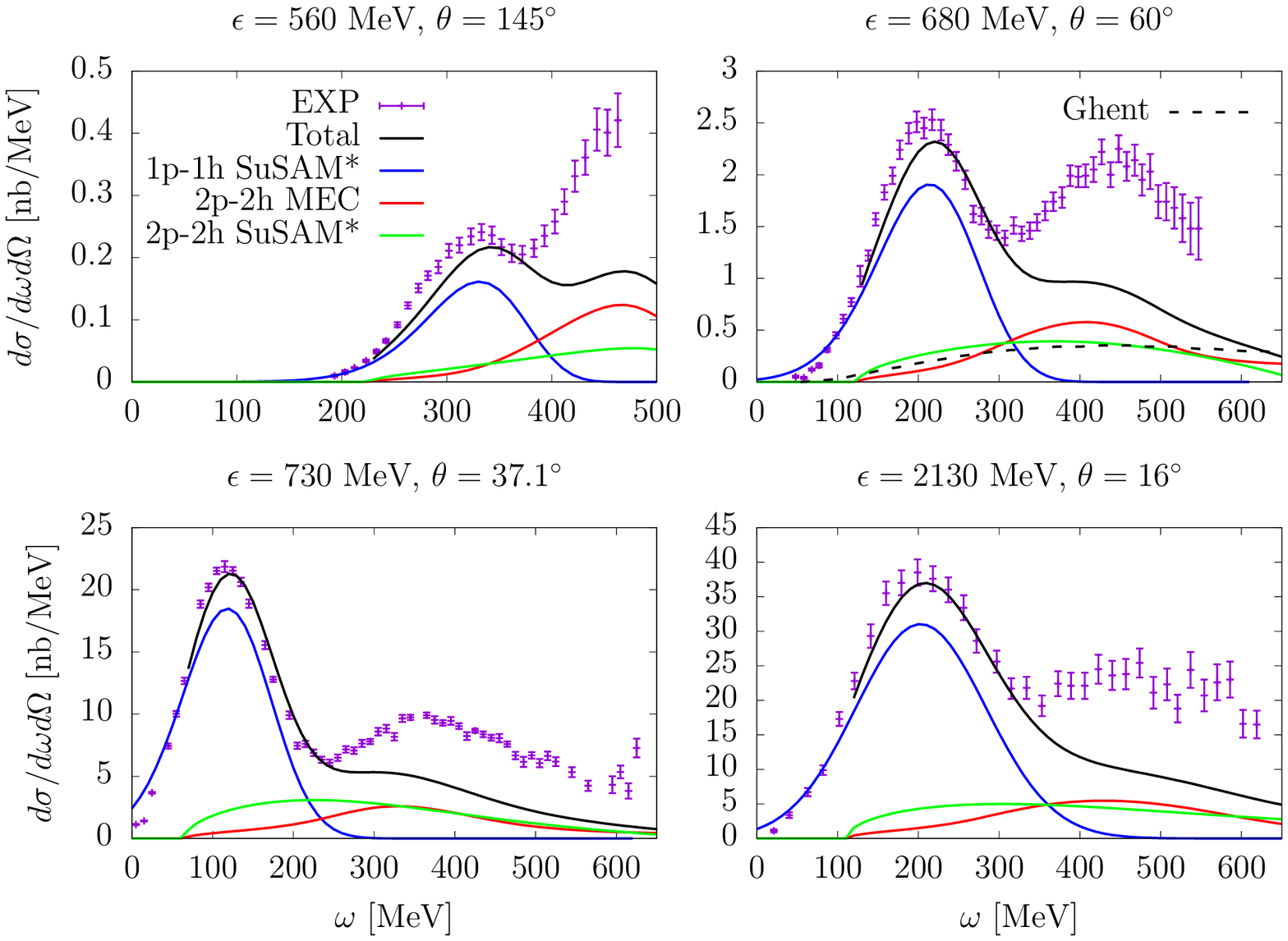}
  \caption{ Calculations of the $^{12}$C$(e,e')$ cross section in the
    separated 1p1h, 2p2h, and MEC channels for several
    kinematics. The total is the sum of the three channels. In the
    case $\epsilon=680$ MeV we compare with the SRC-2p2h of
    Ref. \cite{Cuy16}. The experimental data are from
    \cite{archive,archive2}.  }
\end{figure*}

\begin{figure*}
  \centering
\includegraphics[scale=0.7,bb=40 335 500 670]{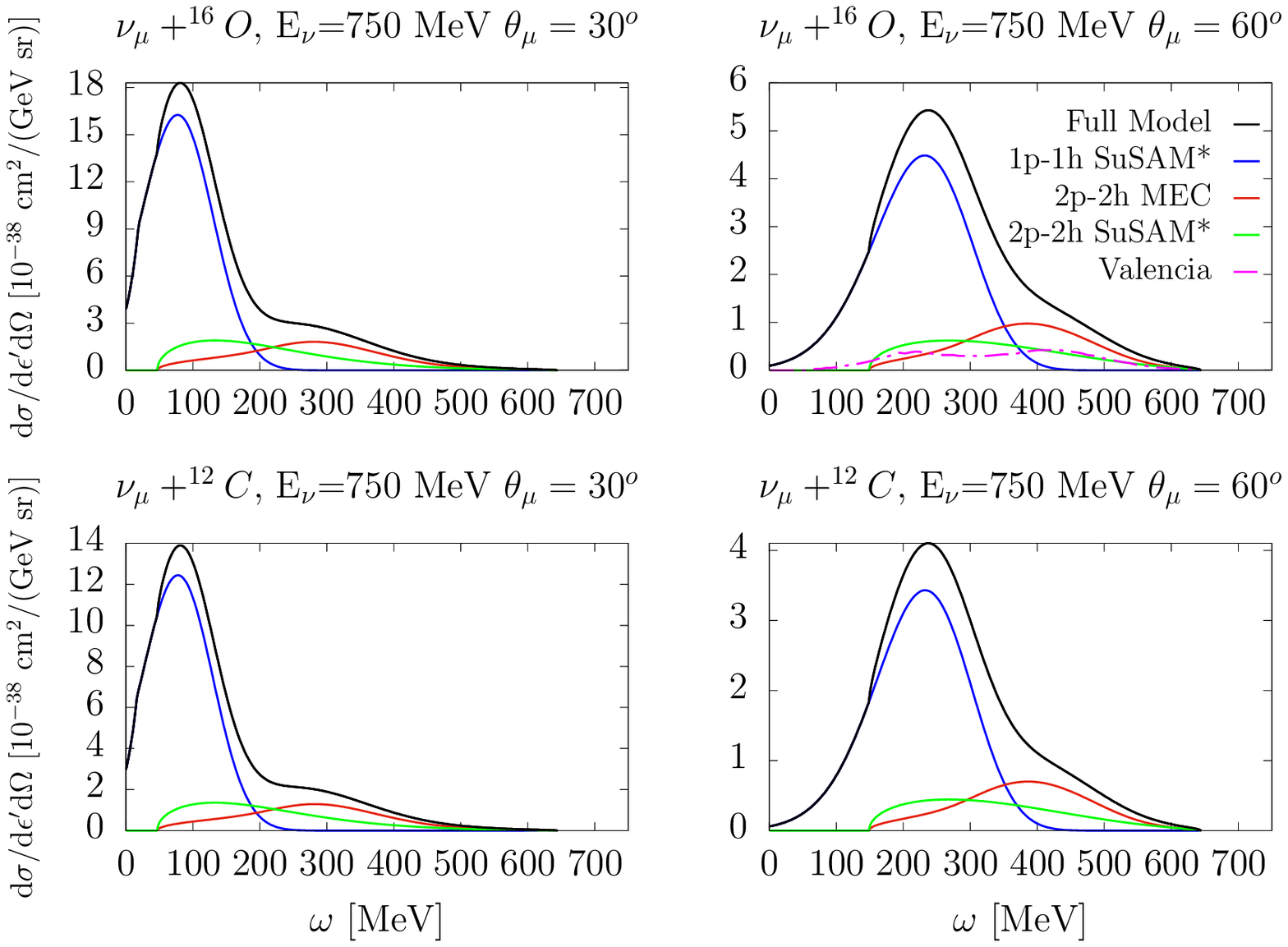}
  \caption{ Calculations of the $^{12}$C and $^{16}$O
    $(\nu_{\mu},\mu)$ cross section in the separated 1p1h, 2p2h,
    and MEC channels for fixed neutrino energy.  The total is the
    sum of the three channels. In the case $^{16}$O,
    $\theta_{\mu}=60^{\rm o}$ we compare with the 2p2h model of
    Ref. \cite{Nie11} }
\end{figure*}

In Fig. 2 we show the 2p2h contribution to the scaling function using
the semiempirical formula, for different values of $q\leq 2000$ MeV/c. 
In all panels the
SuSAM* scaling function, $f_{QE}^*(\psi^*)$, is also shown.
The  2p2h results have been obtained by dividing the
semiempirical formula, Eq. (\ref{semi1}) by the averaged cross
section of the single nucleon as in Eq. (\ref{fexp})
\begin{equation}
f_{2p2h}^* \equiv
\frac{(\frac{d\sigma}{d\Omega d\omega})_{2p2h}}{\sigma_M ( v_L r_L + v_T
  r_T) }.
\label{fsrc}
\end{equation}
Dividing Eq. (\ref{semi1}) over (\ref{rsn}) this gives
\begin{equation}
\label{fsrc2}
f^*_{\rm 2p2h}(q,\omega) =  
\frac{VF(q,\omega) }{(2 \pi)^9} 
\frac{m_N^* \eta_F^3 \kappa}{Z\xi_F} 
\frac{Z+\alpha(Z-1)}{2Z-1}
\frac{ c^{pn}(q)}{m_N^2 m_\pi^4}.
\end{equation}
Note first that $f^*_{\rm 2p2h}(q,\omega)$ does not scale, i.e., it
depends on $\psi$ but also on $q$.  Second the dependency on energy of
$f_{2p2h}^*(q,\omega)$ comes solely from phase-space. This is because
the single nucleon cross section has canceled with the denominator in
Eq. (\ref{fsrc}).  Then here the 2p2h contribution to the scaling function
 is parametrized with a function
increasing with energy as the phase space times a parameter that is
$q$-dependent.  The 2p2h parameter $c^{pn}(q)$ is chosen to make
$f_{2p2h}^*$ to coincide with the tail of the scaling function for
$\psi^*=1.5$.  The 2p2h coefficients
 are given in Table 1 for $q=100$ up to 1000 MeV/c.
In the table we provide two sets of coefficients corresponding to the
parameter $\alpha=1/18$ and also $\alpha=0$ (neglecting the pp and nn
contribution).  It is found that the coefficient $c^{pn}(q)$ decreases
with $q$. Roughly it behaves as $c^{pn}(q) \sim 1/q$.

Since the experimental data of the scaling function are distributed in
a thick band, as seen in Fig. 2, the coefficients $c^{pn}(q)$ have an
uncertainty that can be estimated by fitting the lower or upper part
of the band for $\psi^*=1.5$. The error in $c^{pn}(q)$ 
is approximately 38\% and it is large because
it corresponds to the uncertainty of the scaling function in the tail
zone for $\psi^*=1.5$.

Our fit has been made for $q\leq 1000$ MeV/c that is the region of interest
for electron and neutrino scattering.  For higher $q$ the right side
of the QE peak is missing from the data due to pion emission. 
In fig. 3 we show the contribution
$f^*_{2p2h}$ calculated with Eq (\ref{fsrc2}) for the  kinematics of
all the QE experimental data with $q<1000$ MeV/c.  We see that
these points generate a band that can explain the tail of the scaling
function. To compute $f^*_{\rm 2p2h}$ for arbitrary $q$-values 
we have interpolated the
coefficients $c^{pn}(q)$ using the formula
\begin{equation} \label{interpol}
c^{pn}(q)=a_0\sqrt{\frac{m_N^*+q}{q}} \frac{m_N^*}{q}
\end{equation}
that fits well the $q$ dependence of $c^{pn}(q)$ with the parameter
$a_0=345\pm 130$, for $\alpha=1/18$ (see Fig. 4).

As an example, 
the decrease of $c^{pn}(q)$ is also expected in the particular case of the 
independent-pair factorized model.  
In fact it can be observed from Eq. (28) that the high-momentum function
$\Delta\varphi_{\nh_1\nh_2}(\np)$ 
has a denominator of $h^2-p^2$, causing the average distribution 
$|\Delta\varphi_{\nh_1\nh_2}(\np+\nq/2)|^2$ 
 to rapidly diminish as $q$ increases. 
However in this simplified model, the precise dependence
 may not necessarily follow the exact form described by
equation (49).

Once the coefficients of the semiempirical formula have been fitted, we
propose to carry out a new scaling analysis without the contribution
of the emission of two particles. To do this we subtract the $f_{\rm 2p2h}^*$
contribution from the $f_{QE}^*$  data. Since the MEC contribution had
already been subtracted, the new data no longer contain
2p2h contribution and can thus be considered purely
1p1h data. The result of this new scaling analysis 
is shown in Fig. 5. As we see, after
the subtraction a new band of points is obtained that is 
symmetrical, since the emission of two nucleons has
been eliminated. A new phenomenological scaling function
can now be fitted with a Gaussian 
\begin{equation}\label{f1p1h}
f_{1p1h}^* (\psi^*) = b e^{-(\psi^*)^2 / a^2}.
\end{equation}
The fitted coefficients are a=$0.744\pm 0.082$ and $ b=0.682\pm
0.102$.  The errors in the parameters have been estimated by fitting
Gaussians to the upper and lower part of the band.

With the present approach we can compute the inclusive effective
section of electrons and neutrinos by adding separately the
contributions of SuSAM* 1p1h+2p2h plus MEC contributions.
\begin{equation}\label{stot}
\frac{d \sigma}{d \Omega' d\epsilon'}
=
\left( \frac{d \sigma}{d \Omega' d\epsilon'} \right)_{1p1h}+
\left( \frac{d \sigma}{d \Omega' d\epsilon'} \right)_{\rm 2p2h}+
\left( \frac{d \sigma}{d \Omega' d\epsilon'} \right)_{MEC},
\end{equation}
where the 1p1h cross section is computed with the scaling function 
(\ref{f1p1h}), the 2p2h with the semiempirical formula 
fitted above, and the MEC with the semiempirical formula 
fitted in \cite{Mar21b}.

Results are shown if Fig. 6 for electrons and Fig. 7 for neutrinos.
In Fig. 6 we show the inclusive cross section of $^{12}$C for several
kinematics as a function of $\omega$. We show the separate
contributions of the three terms of Eq. (\ref{stot}) and the total
contribution of the model. The pion emission is not included.
 
The $\omega$-dependence of the SuSAM*-2p2h is similar to the MEC
contribution and extends from the QE peak to the dip region and also
to the $\Delta$ region.  The order of magnitude is similar to the MEC,
except at the $\Delta$-peak where the MEC are larger. For incident
electron energy $\epsilon=680$ MeV we compare with the SRC model of
ref \cite{Cuy16} whose results are quite similar to ours.  The present
results are also similar to the SRC calculation of $^{56}$Fe$(e,e')$
of ref. \cite{Niu22}.

\begin{figure}
  \centering
\includegraphics[scale=0.7,bb=150 330 400 670]{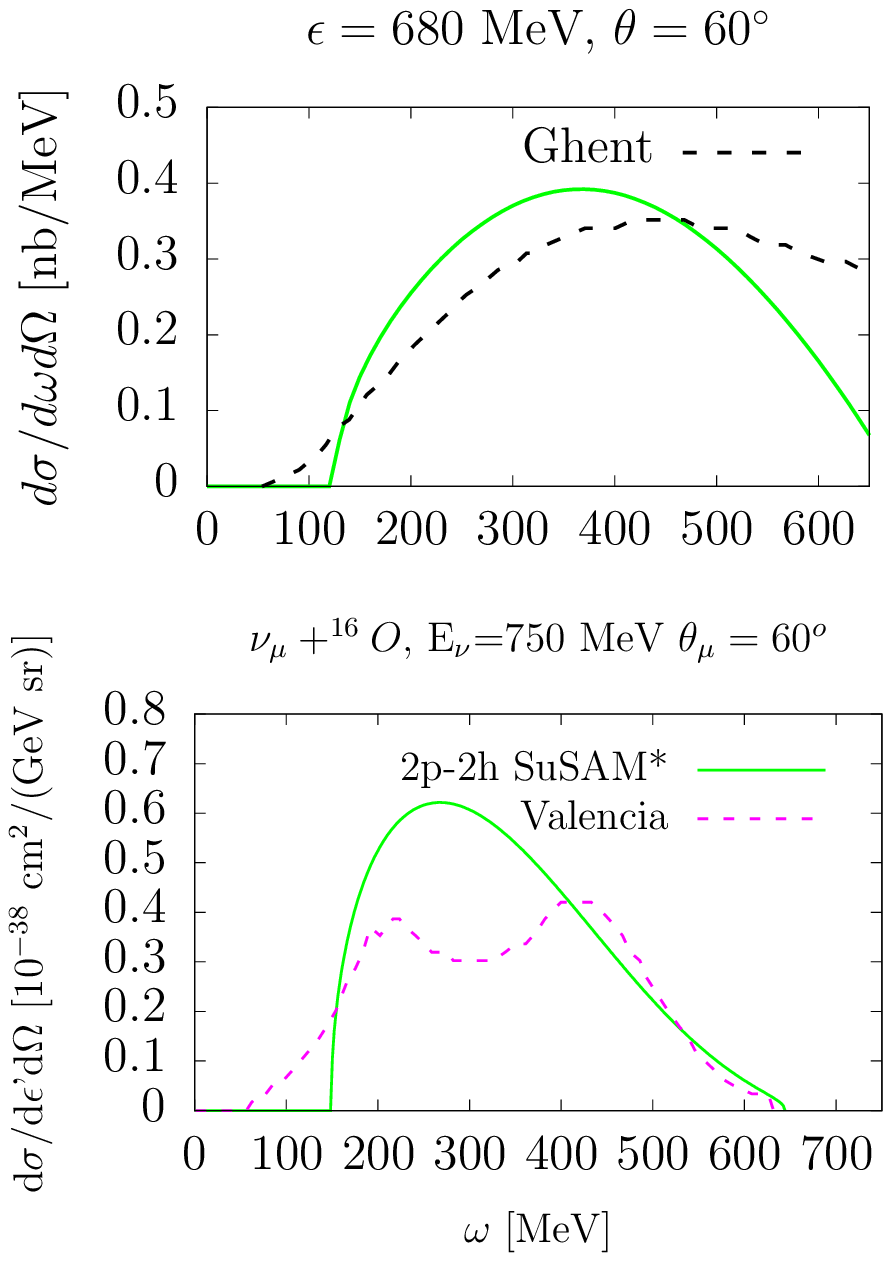}
  \caption{Detailed comparison of the SuSAM*-2p2h results with the 
models of refs. \cite{Cuy16,Nie11} for electron and neutrino scattering.
}
\end{figure}

A similar behavior is seen in Fig. 7 where 
we show the  $(\nu_{\mu},\mu)$ cross section from
$^{16}$O and  
$^{12}$C for fixed neutrino energy and for two scattering angles as a function of $\omega$. A comparison with the SRC results of \cite{Nie11} is also shown
for $^{16}$O and  $\theta_\mu=60^{\rm o}$.

In Fig. 8, we present a detailed comparison between the SuSAM* 2p2h
contribution to the cross section and the SRC models discussed in
references [26] and [13]. Firstly, we observe that both models exhibit
a similar dependence on $\omega$, which is consistent with the 2p2h
phase space upon which our parametrization of the scaling function
tail is based. Additionally, the order of magnitude of the
cross-section is similar to what is obtained with our parametrization
of the 2p2h scaling function. This suggests that both models
reasonably describe the tail of the phenomenological scaling function.

This comparison provides further support for the idea that the tail of
the scaling function can be attributed to 2p2h contributions and that
our parametrization captures its behavior effectively. However, it's
important to note that while these models exhibit qualitative
agreement, there may still be quantitative differences due to various
factors such as different assumptions, formalism, or input parameters
used in each model.

Using the results of these models, we can take advantage of them to
readjust the coefficient $c^{pn}$(q) that results from a microscopic
calculation. In the case of the Ghent calculation, the kinematics
correspond to momentum transfer values varying in a small interval around 
$q=600$ MeV/c. The fit to the semiempirical formula of 2p2h provides a
theoretical value of the coefficient $c^{pn}(q)$, which is shown as a
data point in Figure 4. We see that its value is slightly lower than
the SuSAM* value but within its uncertainty interval.  The kinematics
of the Valencia model corresponds to q values around 650 MeV/c. The
adjusted value of $c^{pn}(q)$ to this model also shown as a data point
in Fig. 4, is also below the SuSAM* value from, but always within the
error interval.

Using the results of these microscopic models to readjust the 2p2h
parameter is a valid approach. While
there are differences between the microscopic models and our
semiempirical formula, the adjusted values of $c^{pn}(q)$ obtained
from these models are consistent with our fits and within the error
intervals.  Indeed, the comparison between different theoretical
calculations, such as those based on electrons and neutrinos and
involving different kinematics, can provide valuable insights into the
compatibility and consistency of these models. The fact that our
phenomenological parameterization of the 2p2h function in the SuSAM*
scaling analysis shows agreement with these theoretical calculations
highlights another useful aspect of the extended scaling approach.

The parametrization of the 2p2h response presented in this scaling
analysis is similar to the pure phase-space model proposed by Mosel et
al. \cite{Mos14}, which is implemented in the GiBUU event generator. In this
scheme, the 2p2h response is parametrized as the product of the 2p2h
phase-space and the dipole electromagnetic form factor, which is
proportional to the single-nucleon response in our
parametrization. The present work further strengthens the validity of this
approach for its use in neutrino scattering cross-section
calculations. Additionally, the parameters of our model are
alternatively fitted to quasielastic scaling data, which is associated
with the high-energy tail of the electron scattering data. This
highlights the potential of this approach to describe both
quasielastic and 2p2h processes in neutrino-nucleus scattering and
supports its application in the analysis of neutrino scattering
experiments.

\section{Conclusions}

In conclusion, this work presents a extension of the superscaling
analysis of electron scattering data using a new parameterization of
the phenomenological scaling function in the quasielastic peak. The
new parameterization assumes that the high-energy tail of the scaling
function is produced by the emission of two nucleons with the one-body
(OB) current through the 2p2h phase space function.  By incorporating
this 2p2h contribution and considering the interference effects with
two-body currents and other processes, we account for the
phenomenological aspects of two-nucleon emission with the OB current
and its contributions to scaling violation.

The new parameterization allows for the restoration of symmetry in the
1p1h scaling function with respect to the scaling variable
$\psi^*$. It effectively separates the 1p1h and 2p2h contributions,
providing a clearer understanding of the underlying physics in the
scaling analysis.  Importantly, the new 1p1h+2p2h parameterization
describes the experimental data as well as the traditional
parameterization. It captures the features of the scaling function,
including the tail region, and provides a consistent description of
the observed scaling phenomena.

We have also studied a simple model based on the independent-pairs
approximation to investigate the basic structure of two-particle
emission in the factorized scaling analysis. In this schematic model,
the 2p2h parameters are related to the average high-momentum
distribution in a 2p2h excitation.  In reality, the coefficients that
characterize the high-energy tail of the scaling function encompass
additional entangled contributions, including interference effects
from MEC and FSI . Although the current phenomenological analysis
offers valuable initial insights, a comprehensive understanding of the
influence of SRCs on the scaling function's high-energy tail
necessitates a more detailed investigation using a realistic model.

An important utility of this extended scaling analysis is that it provides a
procedure to derive a simplified expression for the two-nucleon
emission cross section in nuclei. This simplified expression can be
valuable in reducing the computational effort required for simulations
of neutrino-nucleus interactions. By parametrizing the 2p2h response,
it also offers a method to compare different models of the process,
including comparisons between electron scattering and neutrino
scattering at different kinematics. This capability can be
particularly useful in Monte Carlo event generators used for the
analysis of neutrino oscillation experiments with accelerators.

\section{Acknowledgments}

Work supported by:
Grant PID2020-114767GB-I00
funded by MCIN/AEI/10.13039/501100011033; FEDER/Junta de
Andalucia-Consejeria de Transformacion Economica, Industria,
Conocimiento y Universidades/A-FQM-390-UGR20; and Junta de
Andalucia (Grant No. FQM-225).

\appendix

\begin{widetext}

\section{Matrix element of the OB current with correlated nucleons}
\label{apendiceA}

Here we derive the matrix element of the OB current between a
correlated pair and a two-particle state above the Fermi level,
Eqs. (\ref{matrix},\ref{jsimple}).
It is obtained as the sum of the OB current 
acting on each of the particles.
\begin{eqnarray} 
\langle [\np'_1\np'_2] | J^\mu_{OB}(\nq) | [\Phi_{\nh_1\nh_2}] \rangle =
\langle [\np'_1\np'_2] | J^\mu_{1}+J^\mu_2 | [\Delta\Phi_{\nh_1\nh_2}] \rangle
\end{eqnarray}
Let's calculate the matrix element of $J^\mu_{1}$
writing explicitly the spin indices.
We assume that the two-hole uncorrelated state
is $|\nh_1s_1\nh_2s_2\rangle$. The corresponding 
correlated  wave function is denoted by 
\begin{equation}
|\Phi^{s_1s_2}_{\nh_1\nh_2}\rangle
= |\nh_1s_1\nh_2s_2\rangle
+|\Delta\Phi^{s_1s_2}_{\nh_1\nh_2}\rangle
\end{equation}
where $|\Delta\Phi^{s_1s_2}_{\nh_1\nh_2}\rangle$ carries the high
momentum components.  First we introduce a complete set of momentum
states
\begin{eqnarray} \label{a3}
\langle [\np'_1s'_1\np'_2s'_2] 
| J^\mu_{1}| [\Delta\Phi^{s_1s_2}_{\nh_1\nh_2}] \rangle
= \int d^3 p_1 d^3 p_2 \sum_{\sigma_1\sigma_2}
\langle [\np'_1s'_1\np'_2s'_2] | J^\mu_{1}|\np_1\sigma_1\np_2\sigma_2\rangle
\langle\np_1\sigma_1\np_2\sigma_2| 
[\Delta\Phi^{s_1s_2}_{\nh_1\nh_2}] \rangle
\end{eqnarray}
The first matrix element inside the integral is
\begin{eqnarray}
\langle [\np'_1s'_1\np'_2s'_2] | J^\mu_{1}|\np_1\sigma_1\np_2\sigma_2\rangle
&=&
\langle [\np'_1s'_1] | J^\mu_{1}|[\np_1\sigma_1]\rangle
\langle \np'_2s'_2 | \np_2\sigma_2\rangle
\nonumber\\
&=&
\frac{(2\pi)^3}{V}
\delta(\np'_1-\np_1-\nq)
j^{\mu}_{OB}(\np'_1,\np'_1-\nq)_{s'_1\sigma_1}
\delta(\np'_2-\np_2)\delta_{s'_2\sigma_2}
\label{a4}
\end{eqnarray}
where $j_{OB}$ is the OB current function given in Eq. (\ref{job}) 
in the electromagnetic case, and 
we have exchanged the bracket of $\np'_2$ for $\np_1$.
The matrix element of the high-momentum wave function is 
\begin{eqnarray} 
\langle \np_1\sigma_1\np_2\sigma_2 | [\Delta\Phi^{s_1s_2}_{\nh_1\nh_2}] \rangle
&=& \frac{(2\pi)^3}{V}
\langle \np_1\sigma_1\np_2\sigma_2 | \Delta\Phi^{s_1s_2}_{\nh_1\nh_2} \rangle
\nonumber\\
&=& \frac{(2\pi)^3}{V}
\delta(\np'_1+\np'_2-\nh_1-\nh_2)
\Delta\varphi^{s_1s_2}_{\nh_1\nh_2}(\np)_{\sigma_1\sigma_2}
\label{a5}
\end{eqnarray}
Inserting Eqs. (\ref{a4},\ref{a5}) in (\ref{a3}) and integrating using
the Dirac deltas we obtain
\begin{equation}
\langle [\np'_1s'_1\np'_2s'_2] 
| J^\mu_{1}| [\Delta\Phi^{s_1s_2}_{\nh_1\nh_2}] \rangle 
= \frac{(2\pi)^6}{V^2}
\delta(\np'_1+\np'_2-\nq-\nh_1-\nh_2)
\sum_{\sigma_1}
j^{\mu}_{OB}(\np'_1,\np'_1-\nq)_{s'_1\sigma_1}
\Delta\varphi^{s_1s_2}_{\nh_1\nh_2}
\left(\frac{\np'_1-\np'_2-\nq}{2}\right)_{\sigma_1s'_2}.
\end{equation}
Proceeding analogously to calculate the matrix element of $J_2$, 
we obtain the equation
\begin{equation} 
\langle [\np'_1s'_1\np'_2s'_2] 
| J^\mu_{2}| [\Delta\Phi^{s_1s_2}_{\nh_1\nh_2}] \rangle
= \frac{(2\pi)^6}{V^2}
\delta(\np'_1+\np'_2-\nq-\nh_1-\nh_2)
\sum_{\sigma_2}
j^{\mu}_{OB}(\np'_2,\np'_2-\nq)_{s'_2\sigma_2}
\Delta\varphi^{s_1s_2}_{\nh_1\nh_2}
\left(\frac{\np'_1-\np'_2+\nq}{2}\right)_{s'_1\sigma_2}.
\end{equation}
By analogy with Equation (\ref{e21}), 
the 2p2h correlation current function turns out to be
\begin{eqnarray}
j_{cor}(\np'_1,\np'_2,\nh_1,\nh_2)
&=&
(2\pi)^3
\sum_{\sigma}
j^{\mu}_{OB}(\np'_1,\np'_1-\nq)_{s'_1\sigma}
\Delta\varphi^{s_1s_2}_{\nh_1\nh_2}
\left(\frac{\np'_1-\np'_2-\nq}{2}\right)_{\sigma s'_2}
\nonumber\\
&&
+(2\pi)^3
\sum_{\sigma}
j^{\mu}_{OB}(\np'_2,\np'_2-\nq)_{s'_2\sigma}
\Delta\varphi^{s_1s_2}_{\nh_1\nh_2}
\left(\frac{\np'_1-\np'_2+\nq}{2}\right)_{s'_1\sigma}.
\label{jcor}
\end{eqnarray}
\end{widetext}

\section{Averaged single-nucleon responses in the SuSAM* model}
\label{apendiceB}

The single-nucleon responses can be extended beyond values $|\psi^*|>$1
using the energy distribution function (\ref{MDF}) obtained form the
superscaling function by differentiating the two sides of the equation
(14) with respect to $\epsilon_0$
\begin{eqnarray}\label{MDF}
n(\epsilon_0)= - \frac{2}{3} \frac{1}{\psi^*} \frac{df^*(\psi^*)}{d \psi^*}.
\end{eqnarray}
We use the non-correlated superscaling function
$f^*(\psi^*)=b\exp(-(\psi^*)^2/a^2)$ to compute de energy distribution and the
averaged single-nucleon responses, Eq.  (\ref{def_sn}), giving
$n(\epsilon_0)= -\frac{4}{3a^2}f^*(\psi^*)$.  The single-nucleon responses
$U_K(\epsilon,q,\omega)$ for neutrinos (antineutrinos) are given in
the Appendix C of Ref. \cite{Ama20}. The averaged single-nucleon
responses after integration over $\epsilon$ are given
by
\begin{eqnarray}
\overline{U}_{CC} 
&=& 
- w_1 \frac{\kappa^2}{\tau} + w_2 \lambda^2 +  2 \lambda w_2  \left[ 1+\xi_F(\psi^{*2}+a^2) \right] \nonumber\\
&& 
\kern -1cm \mbox{}
+w_2 \left[ 
1 + 2 \xi_F (\psi^{*2}+ a^2)+ \xi_F^2 (\psi^{*4}+2a^2\psi^{*2}+2a^4)) 
\right] \nonumber \\
&+& 4 \lambda^2w_4, 
\end{eqnarray}
\begin{eqnarray}
\overline{U}_{CL} 
&=& 
\frac{\lambda}{\kappa}
\Bigl\{ 
w_1 \frac{\kappa^2}{\tau} - w_2 \lambda^2 
 - 2 \lambda w_2  \left[ 1+\xi_F(\psi^{*2}+a^2) \right] 
\nonumber\\
&&
\kern -1cm \mbox{} 
- w_2 
\left[ 1 + 2 \xi_F (\psi^{*2}+ a^2) + \xi_F^2 (\psi^{*4}+2a^2\psi^{*2}+2a^4) 
\right] 
\Bigr\}
\nonumber\\
&& 
\mbox{}- 4 \lambda\kappa w_4,
\end{eqnarray}
\begin{eqnarray}
\overline{U}_{LL} 
&=& 
\frac{\lambda^2}{\kappa^2}
\Bigl\{ - w_1 \frac{\kappa^2}{\tau} + w_2 \lambda^2   + 2 \lambda w_2  \left[ 1+\xi_F(\psi^{*2}+a^2) \right]  
\nonumber\\
&&
\kern -1cm \mbox{} 
+ w_2 \left[ 1 + 2 \xi_F (\psi^{*2}+ a^2) + \xi_F^2 (\psi^{*4}+2a^2\psi^{*2}+2a^4) \right] \Bigr\}
\nonumber\\
&+&  4 \kappa^2w_4, 
\end{eqnarray}
\begin{eqnarray}
\overline{U}_{T} 
&=&   
2 w_1 + 2w_2\lambda \frac{\tau}{\kappa^2} [  1+\xi_F(\psi^{*2}+a^2)]-w_2   
\nonumber  \\ 
&& 
\kern -1cm
\mbox{} +
w_2 \frac{\tau}{\kappa^2}  \left[1 + 2 \xi_F (\psi^{*2}+ a^2) + \xi_F^2 (\psi^{*4}+2a^2\psi^{*2}+2a^4)\right] \nonumber\\
&-& \frac{\tau^2}{\kappa^2} w_2, 
\end{eqnarray}
\begin{eqnarray}
\overline{U}_{T'} = 2 w_3\frac{\tau}{\kappa} \left\lbrace\lambda +
[1+\xi_F(\psi^{*2}+a^2)]\right \rbrace.
\end{eqnarray}
The single-nucleon structure functions, $w_1$, $w_2$, $w_3$ and $w_4$
are given \cite{Ama20} by
\begin{eqnarray}
w_1 &=& \tau (2G_M^{*V})^2 + (1+\tau) G_A^2, \\
w_2 &=& \frac{(2G_E^{*V})^2 +\tau (2G_M^{*V})^2}{1+\tau} + G_A^2, \\
w_3&=& G_A 2G_M^{*V}, \\
w_4&=& \frac{(G_A-\tau G^*_P)^2}{4\tau},
\end{eqnarray}
where $G_E^{*V}$ and $G_M^{*V}$ are the electric and magnetic isovector 
form factors $G_{E,M}^{*V}=(G_{E,M}^{*P}-G_{E,M}^{*N})/2$
modified in the medium with the relativistic effective mass 
\cite{Ama15}.
For the axial form factor, $G_A$, 
we use dipole parametrization with axial mass $M_A=1.032$ GeV.
Finally the pseudoscalar form factor $G^*_P=4m_Nm_N^* G_A/(m_\pi^2-Q^2)$.

For electron scattering the equations are similar for
$\overline{U}_L=\overline{U}_{CC}$ and $\overline{U}_T$, without
the axial contribution, $w_3=w_4=0$. 
For protons and neutrons the corresponding structure functions are
\begin{eqnarray}
w^{em}_1 &=& \tau (G_M^{*})^2, \\
w^{em}_2 &=& \frac{(G_E^{*})^2 +\tau (G_M^{*})^2}{1+\tau},  
\end{eqnarray}
with the electric and magnetic form factors
\begin{equation}
G_E^*  =  F_1-\tau \frac{m^*_N}{m_N} F_2, \kern 1cm
G_M^*  = F_1+\frac{m_N^*}{m_N} F_2.  \label{GM}
\end{equation}
For the $F_i$ form factors of the nucleon, we use the Galster
parameterization \cite{Gal71}.



\begin{thebibliography}{99}
\bibitem{Ank22}
A.~M.~Ankowski, A.~Ashkenazi, S.~Bacca, J.~L.~Barrow, M.~Betancourt, A.~Bodek, M.~E.~Christy, L.~D.~S.~Dytman, A.~Friedland and O.~Hen, \textit{et al.}
[arXiv:2203.06853 [hep-ex]].

\bibitem{Ama20}
J.~E.~Amaro, M.~B.~Barbaro, J.~A.~Caballero, R.~Gonz\'alez-Jim\'enez, G.~D.~Megias and I.~Ruiz Simo,
J. Phys. G \textbf{47} (2020) no.12, 124001

\bibitem{Mos16} U. Mosel, Ann. Rev. Nuc. Part. Sci. 66 (2016), 171.

\bibitem{Kat17}  T.~Katori and M.~Martini,
  J.\ Phys.\ G {\bf 45} (2018) no.1,  013001.

\bibitem{Alv14} 
L. Alvarez-Ruso, Y. Hayato, J. Nieves, New J. Phys. 16 (2014) 075015.

\bibitem{Ank17}  A. M. Ankowski, C. Mariani, J. Phys. G44 (2017) 054001.

\bibitem{Ben17}
 O. Benhar, P. Huber, C. Mariani, D. Meloni, Phys. Rep. 700 (2017) 1.

\bibitem{Ros80} R. Rosenfelder, Ann. Phys. (N.Y.) 128, 188 (1980).

\bibitem{Ser86} B.D. Serot, and J.D. Walecka, Adv. Nucl. Phys. 16 (1986) 1.

\bibitem{Dre89} D Drechselt and M M Giannini,
Rep. Prog. Phys. 52 (1989) 1083.

\bibitem{Weh93} K. Wehrberger, Phys. Rep. 225 (1993) 273.


\bibitem{Mar09} M. Martini, M. Ericson, G. Chanfray, J. Marteau, 
                 Phys.Rev. C80 (2009) 065501.

\bibitem{Nie11} J. Nieves, I. Ruiz Simo, M.J. Vicente Vacas, 
Phys.Rev. C83 (2011) 045501.

\bibitem{Gal16} 
  K.~Gallmeister, U.~Mosel and J.~Weil,
  Phys.\ Rev.\ C {\bf 94}, no. 3, 035502 (2016).

\bibitem{Meg14} 
  G.~D.~Megias, M.~V.~Ivanov, R.~Gonzalez-Jimenez, M.~B.~Barbaro, J.~A.~Caballero, T.~W.~Donnelly and J.~M.~Udias,
  Phys.\ Rev.\ D {\bf 89}, no. 9, 093002 (2014)
  Erratum: [Phys.\ Rev.\ D {\bf 91}, no. 3, 039903 (2015)]


\bibitem{Meg16}
 G.D Megias, 
J.~E.~Amaro, M.~B.~Barbaro, J.~A.~Caballero, T.~W.~Donnelly and I. Ruiz Simo,
 Phys. Rev. D 94 (2016), 093004. 






\bibitem{Meg16b} 
  G.~D.~Megias, J.~E.~Amaro, M.~B.~Barbaro, J.~A.~Caballero and T.~W.~Donnelly,
  Phys.\ Rev.\ D {\bf 94}, 013012 (2016).

\bibitem{Ank15} 
  A.~M.~Ankowski,
  Phys.\ Rev.\ D {\bf 92}, no. 1, 013007 (2015).

\bibitem{Gra13} 
  R.~Gran, J.~Nieves, F.~Sanchez and M.~J.~Vicente Vacas,
  Phys.\ Rev.\ D {\bf 88}, no. 11, 113007 (2013).

\bibitem{Pan16} 
  V.~Pandey, N.~Jachowicz, M.~Martini, R.~Gonzalez-Jimenez, J.~Ryckebusch, T.~Van Cuyck and N.~Van Dessel,
  Phys.\ Rev.\ C {\bf 94}, no. 5, 054609 (2016)

\bibitem{Mar16} 
  M.~Martini, N.~Jachowicz, M.~Ericson, V.~Pandey, T.~Van Cuyck and N.~Van Dessel,
  Phys.\ Rev.\ C {\bf 94}, no. 1, 015501 (2016).


\bibitem{Mar21}
V.~L.~Martinez-Consentino, I.~R.~Simo and J.~E.~Amaro,
Phys. Rev. C \textbf{104}, no.2, 025501 (2021)

\bibitem{Mar21b}
V.~L.~Martinez-Consentino, J.~E.~Amaro and I.~Ruiz Simo,
Phys. Rev. D \textbf{104}, no.11, 113006 (2021)




\bibitem{Pac03}
A.~De Pace, M.~Nardi, W.~M.~Alberico, T.~W.~Donnelly and A.~Molinari,
Nucl. Phys. A \textbf{726}, 303-326 (2003)

\bibitem{Rui17}
I.~Ruiz Simo, J.~E.~Amaro, M.~B.~Barbaro, A.~De Pace, J.~A.~Caballero and T.~W.~Donnelly,
J. Phys. G \textbf{44}, no.6, 065105 (2017)


\bibitem{Cuy16}
T.~Van Cuyck, N.~Jachowicz, R.~Gonz\'alez-Jim\'enez, M.~Martini, V.~Pandey, J.~Ryckebusch and N.~Van Dessel,
Phys. Rev. C \textbf{94}, no.2, 024611 (2016)

\bibitem{Ryc97}
J.~Ryckebusch, V.~Van der Sluys, K.~Heyde, H.~Holvoet, W.~Van Nespen, M.~Waroquier and M.~Vanderhaeghen,
Nucl. Phys. A \textbf{624}, 581-622 (1997)

\bibitem{Ste18}
S.~Stevens, J.~Ryckebusch, W.~Cosyn and A.~Waets,
Phys. Lett. B \textbf{777}, 374-380 (2018)


\bibitem{Cos21}
W.~Cosyn and J.~Ryckebusch,
Phys. Lett. B \textbf{820}, 136526 (2021)

\bibitem{Alb84} W. Alberico, M. Ericson, and A. Molinari, Ann. Phys. (N.Y.) 154 (1984) 356.

\bibitem{Alb90}
W.~M.~Alberico, T.~W.~Donnelly and A.~Molinari,
Nucl. Phys. A \textbf{512}, 541-590 (1990)

\bibitem{Ama02}
J.~E.~Amaro, M.~B.~Barbaro, J.~A.~Caballero, T.~W.~Donnelly and A.~Molinari,
Phys. Rept. \textbf{368}, 317-407 (2002)


\bibitem{Ama10}
J.~E.~Amaro, C.~Maieron, M.~B.~Barbaro, J.~A.~Caballero and T.~W.~Donnelly,
Phys. Rev. C \textbf{82}, 044601 (2010)



\bibitem{Ryc19}
J.~Ryckebusch, W.~Cosyn, S.~Stevens, C.~Casert and J.~Nys,
Phys. Lett. B \textbf{792}, 21-28 (2019)

\bibitem{Sim17}
I.~Ruiz Simo, R.~Navarro P\'erez, J.~E.~Amaro and E.~Ruiz Arriola,
Phys. Rev. C \textbf{96}, no.5, 054006 (2017)

\bibitem{Col15}
C.~Colle, O.~Hen, W.~Cosyn, I.~Korover, E.~Piasetzky, J.~Ryckebusch and L.~B.~Weinstein,
Phys. Rev. C \textbf{92}, no.2, 024604 (2015)


\bibitem{Wei21}
R.~Weiss, A.~W.~Denniston, J.~R.~Pybus, O.~Hen, E.~Piasetzky, A.~Schmidt, L.~B.~Weinstein and N.~Barnea,
Phys. Rev. C \textbf{103}, no.3, L031301 (2021)


\bibitem{Ngu20}
D.~Nguyen \textit{et al.} [Jefferson Lab Hall A],
Phys. Rev. C \textbf{102}, no.6, 064004 (2020)

\bibitem{Ama17} J.~E.~Amaro,  I.~Ruiz Simo,  and E.~Ruiz Arriola,   
  Phys.\ Rev.\ D {\bf 95}, 076009 (2017).

\bibitem{Mar17} 
  V.~L.~Martinez-Consentino, I.~Ruiz Simo, J.~E.~Amaro and E.~Ruiz Arriola,
  Phys.\ Rev.\ C {\bf 96}, no. 6, 064612 (2017).

\bibitem{Ama18}
J.~E.~Amaro, V.~L.~Martinez-Consentino, E.~Ruiz Arriola and I.~Ruiz Simo,
Phys. Rev. C \textbf{98} (2018) 024627

\bibitem{Rui18} I.~Ruiz Simo,   V.~L.~Martinez-Consentino, 
                J.~E.~Amaro and E.~Ruiz Arriola,
                Phys.\ Rev.\ D {\bf 97}, 116006 (2018).


MDPI and ACS Style
\bibitem{Cas23}
P.R. Casale, J.E.  Amaro, V.L. Martinez-Consentino, I. Ruiz Simo,
Universe {\bf 9} (2023) 158. 

\bibitem{Sar93}
A. M. Saruis,
Phys. Rept. \textbf{235}, 57-188 (1993)

\bibitem{Wes75}
G.~B.~West,
Phys. Rept. \textbf{18}, 263-323 (1975)

\bibitem{Wal95}
J.D. Walecka, Theoretical Nuclear and Subnuclear Physics,
Oxford University Press (New York) 1995.

\bibitem{Rui17b}
I.~Ruiz Simo, R.~Navarro P\'erez, J.~E.~Amaro and E.~Ruiz Arriola,
Phys. Rev. C \textbf{96}, no.5, 054006 (2017)


\bibitem{Nav13}
R.~Navarro P\'erez, J.~E.~Amaro and E.~Ruiz Arriola,
Phys. Rev. C \textbf{88}, no.6, 064002 (2013)
[erratum: Phys. Rev. C \textbf{91}, no.2, 029901 (2015)]



\bibitem{Jef07}
R.~Shneor \textit{et al.} [Jefferson Lab Hall A],
Phys. Rev. Lett. \textbf{99}, 072501 (2007)


\bibitem{Sub08}
R.~Subedi, R.~Shneor, P.~Monaghan, B.~D.~Anderson, K.~Aniol, J.~Annand, J.~Arrington, H.~Benaoum, W.~Bertozzi and F.~Benmokhtar, \textit{et al.}
Science \textbf{320}, 1476-1478 (2008)


\bibitem{archive} O. Benhar, D. Day and I. Sick, arXiv:nucl-ex/0603032.

\bibitem{archive2} O. Benhar, D. Day, and I. Sick, 
http://faculty.virginia.edu/qes-archive/  

\bibitem{Sim14b}
I.~Ruiz Simo, C.~Albertus, J.~E.~Amaro, M.~B.~Barbaro, J.~A.~Caballero and T.~W.~Donnelly,
Phys. Rev. D \textbf{90}, no.5, 053010 (2014)

\bibitem{Sim14}
I.~Ruiz Simo, C.~Albertus, J.~E.~Amaro, M.~B.~Barbaro, J.~A.~Caballero and T.~W.~Donnelly,
Phys. Rev. D \textbf{90}, no.3, 033012 (2014)

\bibitem{Sim17b}
I.~Ruiz Simo, J.~E.~Amaro, M.~B.~Barbaro, J.~A.~Caballero, G.~D.~Megias and T.~W.~Donnelly,
Phys. Lett. B \textbf{770}, 193-199 (2017)

\bibitem{Co01}
G.~Co' and A.~M.~Lallena,
Annals Phys. \textbf{287}, 101-150 (2001)

\bibitem{Co98}
G.~Co' and A.~M.~Lallena,
Phys. Rev. C \textbf{57}, 145-149 (1998)


\bibitem{Niu22}
  Q. Niu, J. Liu, Y. Guo, C.  Xu, M.  Lyu, and Z. Ren,
 PhysRev C 105 (2022) L051602.

\bibitem{Ama15} 
  J.~E.~Amaro, E.~Ruiz Arriola and I.~Ruiz Simo,
  Phys.\ Rev.\ C {\bf 92}, no. 5, 054607 (2015).




\bibitem{Gal71} 
  S.~Galster, H.~Klein, J.~Moritz, K.~H.~Schmidt, D.~Wegener and J.~Bleckwenn,
  Nucl.\ Phys.\ B {\bf 32}, 221 (1971).



\bibitem{Mos14}
U.~Mosel, O.~Lalakulich and K.~Gallmeister,
Phys. Rev. D \textbf{89}, no.9, 093003 (2014)



\end{thebibliography}
\end{document}